\documentclass[12pt,preprint]{aastex}

\newcommand\chb{$C({\rm H}\beta)$}
\newcommand\object{}
\newcommand{\grsim}{\mathrel{\hbox{\rlap{\hbox{\lower4pt\hbox{$\sim$}}}\hbox{$>$}}}}

\shorttitle{The Primordial Helium Abundance}
\shortauthors{Luridiana et al.}

\begin{document}

\title{The effect of collisional enhancement of Balmer lines\\
        on the determination of the primordial helium abundance}

\author{V. Luridiana\altaffilmark{1}}
\affil{Instituto de Astrof\'\i sica de Andaluc\'\i a (CSIC), Granada, Spain}
\author{A. Peimbert\altaffilmark{2} and M. Peimbert\altaffilmark{2}}
\affil{Instituto de Astronom\'\i a (UNAM), M\'exico D.F., Mexico}
\and
\author{M. Cervi\~no\altaffilmark{1,3}}
\affil{Instituto de Astrof\'\i sica de Andaluc\'\i a (CSIC), Granada, Spain}
\affil{Laboratorio de Astrof\'\i sica Espacial y F\'\i sica Fundamental (INTA),
Madrid, Spain}
\email{vale@iaa.es, antonio@astroscu.unam.mx, peimbert@astroscu.unam.mx, mcs@laeff.esa.es}

\begin{abstract}
This paper describes a new determination of the 
primordial helium abundance ($Y_P$),
based on the abundance analysis of 
five metal-poor extragalactic H{\sc~ii} regions. 
For three regions of the sample
(\object{SBS~0335--052}, \object{I~Zw~18}, and \object{H~29})
we present tailored photoionization models 
based on improved calculations with respect to previous models.
In particular, we use the photoionization models
to study quantitatively the effect of 
collisional excitation of Balmer lines on the determination of 
the helium abundance ($Y$) in the individual regions.
This effect is twofold: 
first, the intensities of the Balmer lines are enhanced 
with respect to the pure recombination value, 
mimicking a higher hydrogen abundance; 
second, the observed reddening is larger
than the true extinction, due to the differential
effect of collisions on different Balmer lines.
In addition to these effects,
our analysis takes into account the following features
of H{\sc~ii} regions:
(i) the temperature structure,
(ii) the density structure,
(iii) the presence of neutral helium,
(iv) the collisional excitation of the He{\sc~i} lines, 
(v) the underlying absorption of the He{\sc~i} lines, and
(vi) the optical thickness of the He{\sc~i} lines.
The object that shows the highest increase in $Y$ after 
the inclusion of collisional effects in the analysis
is \object{SBS~0335--052},
whose helium abundance has been revised by
$\Delta Y = +0.0107$.
The revised $Y$ values for the five objects 
in our sample yield an increase of $+0.0035$ in $Y_P$,
giving $Y_P = 0.2391 \pm 0.0020$.
\end{abstract}

\keywords{galaxies: abundances --- galaxies: individual (\object{SBS~0335--052}, \object{I~Zw~18}, \object{H~29}) --- galaxies: ISM --- H{\sc~ii} regions --- ISM: abundances}

\section{Introduction}

In this work we describe a new determination of the 
primordial $^4$He abundance ($Y_P$),
based on the abundance analysis of 
five metal-poor extragalactic H{\sc~ii} regions. 
The value of $Y_P$ is still one of the missing pieces
in the cosmological scenario.
According to the standard model of Big-Bang nucleosynthesis, 
the primordial abundance of four light isotopes
(D, $^7$Li, $^3$He, and $^4$He)
depends on one parameter only, the baryon-to-photon ratio.
Therefore, the four abundance values
provide altogether a strong constraint on the cosmological models.
Unfortunately, their determination is not an easy task,
and each of the four isotopes poses a different challenge.
In particular, although $^4$He is the most abundant of the four
and the easiest to measure,
it is also the less sensitive to the baryon-to-photon fraction:
this feature implies that $Y_P$ determinations must be both
accurate to be truthful, and precise to be meaningful. 
Current $Y_P$ determinations are thought to be accurate to the third decimal digit,
a level at which differences still matter
from the point of view of cosmology.
To improve further, 
one must take into account all the sources of uncertainty
that affect, down to this level, both the accuracy and the precision of $Y_P$ determinations.
In the following, we will briefly discuss some of these sources;
see also the reviews by \citet{L03} and
\citet{P03m} for a more complete discussion of this topic 
and a quantitative estimate of the error budget in the determination of $Y_P$.

According to the standard scenario, 
the universe was born with zero metallicity ($Z$);
hence, $Y_P$ can be determined extrapolating to $Z=0$ 
the relationship between $Z$ and
the $^4$He abundance ($Y$) for a sample of objects.
This procedure relies on the determination 
of the individual $Y$ and $Z$ values, 
and of the slope $dY/dZ$ of the $Y$ vs. $Z$ curve, which is assumed linear.
The uncertainty affecting $Y_P$ depends directly on 
the uncertainties affecting either
of these basic ingredients, i.e. $dY/dZ$ and the ensemble of the ($Y$, $Z$) pairs.
For this reason, it has long been thought that the best
results are obtained 
from the analysis of extremely low-metallicity objects, such as
dwarf irregulars (dIrrs) and blue compact dwarf galaxies (BCDs),
since the use of these objects minimizes the
uncertainty associated to $dY/dZ$.
However, \citet{P03a} and \citet{P03m} have noted that
this advantage is outweighed by the relatively
higher uncertainty on the $Y$ values,
which derives from the (unknown) 
collisional contribution to the Balmer line intensities:
an uncertainty especially affecting these objects
since  collisional contribution is quite important at high temperatures,
and rapidly fades away at intermediate and low temperatures.
As we will see throughout the paper, a precise estimate
of this contribution is rather difficult to obtain,
and the resulting uncertainty more than offsets 
the smaller uncertainty in the extrapolation to $Z=0$.
\citet{P03a} suggests to use moderately low-metallicity regions
in the determination of $Y_P$,
because in those regions the collisional contribution can be neglected,
and the effect of the uncertainty on the slope is still moderate.

This paper is concerned mainly with the estimation of the collisional
contribution to the Balmer lines in low-metallicity regions,
and of both the correction 
and the additional uncertainty it introduces in the value of $Y_P$.
In a previous paper \citep*[][hereinafter Paper~I]{PPL02} 
we discussed the effect of temperature 
structure on the determination of $Y_P$, 
and derived a new value of $Y_P$ based on a sample
of five extragalactic H{\sc~ii} regions with low and very low metallicities:
\object{NGC~346}, \object{NGC~2363}, \object{H~29}, \object{SBS~0335--052}, 
and \object{I~Zw~18};
Paper~I also contains a preliminary discussion
of the effect of collisions on the determination of $Y_P$.
The abundance analysis of these objects is based on the combined use of 
standard empirical relations and tailored photoionization
models.
In the case of \object{NGC 2363}, we used the models described
in \citet*{LPL99}, 
while \object{NGC 346} has been modeled by \citet*{RPB02}.
The models for the remaining three objects 
will be described in the present paper.

\section{The models}

The models presented in this paper are based on improved
calculations with respect to those mentioned in Paper I.
The main changes introduced are the following:
(i) the photoionization code Cloudy 94 \citep{F00a,F00b} 
    was used instead of the older version Cloudy 90
    (an exhaustive list of the differences between 
    the two versions can be found at {\tt http://www.pa.uky.edu/$\sim$gary/cloudy});
(ii) the geometry of the models was modified to improve
the fit with the observed quantities;
(iii) the ionizing sources were computed with the population 
synthesis code described in \citet{CVGal02},
instead of Starburst99 \citep{Lal99}.

\subsection{Ionizing spectra}

The synthesis code by \citet{CVGal02} allows taking into account
the sampling effects in the initial mass function (IMF),
which is filled stochastically following a Monte Carlo method.
While in very massive stellar populations the IMF can be effectively
described by an analytical law, in smaller populations
sampling effects can be important, and an analytical description of the IMF
artificially introduces fractional numbers of stars in the high-mass end
of the mass spectrum, where the statistics is intrinsically low.
As a result, those parts of the spectrum depending strongly on
the massive star population may be unrealistically represented
by an analytical description.
The sense of using Monte Carlo simulations is therefore the search for self-consistency,
rather than the goal of reproducing the specific stellar content of these clusters.
Stated otherwise, we don't aim at reproducing the observed cluster as it is,
since specific information on the stellar content of the regions analyzed
is not available;   
we rather attempt to build a synthesis model that is both physically
self-consistent, and capable of reproducing the observational constraints.
This conditions are sufficient to build meaningful photoionization models.

All the models described in this work assume a Salpeter IMF in the mass range 2 -- 120
M$_\odot$, and evolutionary tracks with stellar metallicity 
$Z_*=0.001$ and the standard mass-loss rates by \cite{Sal92a};
the stellar metallicity is the lowest available value
among the stellar libraries included in the synthesis code.  
As for the stellar atmospheres,
the code uses the models by \citet{SK97} 
at $Z=0.004$ for main sequence (MS) hot stars, 
those by \citet{Sal92b} for Wolf-Rayet (WR) stars, 
and those by \cite{K91} at $Z$ = 0.002 for the remaining stars.  
As a safety check, we have compared the resulting effective rates of ionizing
photons \citep{Cal03} for H$^0$, He$^0$, He$^+$, O$^0$, O$^+$, and O$^{++}$ 
to those obtained with the new version of Starburst99 
with the atmosphere models by \citet{Sal02} and 
evolutionary tracks with high mass-loss rates \citep{Mal94}. 
The resulting effective rates of ionizing photons in the two codes
for H$^0$, He$^0$, O$^0$, and O$^+$ coincide to better
than 0.15 dex for the age range considered in this work, 
despite the differences in the evolutionary tracks and the atmosphere models. 
On the other hand, there are strong differences, amounting to more than 0.9 dex, 
for the effective ionizing rates of He$^+$ and O$^{++}$,
which determine the abundances of  He$^{++}$ and O$^{3+}$ respectively. 
Hence, our results are virtually independent on the model atmospheres 
and evolutionary tracks assumed, 
except for the case of He$^{++}$ (which will be not be used as a 
constraint: see Section~\ref{sec:obs_constraints} below),
and the case of O$^{3+}$, which is an unimportant ion.

The spectra used in the modeling were computed 
using the following procedure: 
(i) from
the emission rate of ionizing photons $Q({\rm H}^0)$ inferred from the observations,
and through a comparison with the results of a completely sampled IMF,
we have obtained a first estimate of the mass transformed
into stars and of the number of stars 
for the given IMF and mass limits;
(ii) from the observed features of the spectrum,
we have constrained the age range of each object;
for example, the presence of WR features 
implies that the age of the starburst is in the range 2.9 -- 3.1 Myr 
for the case of an instantaneous burst (IB),
or larger than 2.9 Myr for the case of a continuous star formation (CSF) burst;
(iii) with the input parameters obtained in the previous steps,
we have performed a large number of Monte Carlo simulations;
(iv) among all the simulations, we have selected 
those with a $Q({\rm H}^0)$ value consistent with the observations;
(v) for those H{\sc~ii} regions whose spectra show WR features
we have further selected the simulations including WR stars;
(vi) finally, we have chosen the simulations with the largest value 
of the ratio $Q({\rm O}^{+})/Q({\rm H}^0)$,
where $Q({\rm O}^{+})$ is the emission rate of photons
with energy $E > 2.6$ Ryd;
this choice is motivated by the general tendency 
in most of our trial models of emitting  globally too low oxygen lines:
this criterion is a simple way of overcoming this difficulty,
since a harder spectra increases the $R_{23}$ parameter
\citep{Pal79}.
The particular features of each object will be discussed
in the corresponding section.

\subsection{Geometry}

This section describes the geometric assumptions 
made in our models.
One of the most important features of H{\sc~ii} regions 
affecting the geometric assumptions made in the modeling
is photon leak:
for example, \citet{ZRB01} estimate that 25 to 60 percent of the
ionizing photons escape from giant H{\sc~ii} regions.
In a spherically symmetric model,
the escape of photons can be modeled in two basic ways,
which correspond to two different geometrical scenarios.
The first assumes that the actual
radius is smaller than the Str\"omgren radius ($R_S$), or, equivalently, that
the gas cloud ends before it succeeds in absorbing all the ionizing photons.
A model with this feature is {\it density bounded}
(the radius is determined by the quantity of gas available),
whereas a model reaching the Str\"omgren radius is 
{\it radiation bounded} (the radius is determined by the
rate of ionizing photons emitted by the central source).
A density bounded model can be roughly thought of as a radiation bounded model
from which the outer zones have been stripped away: since these zones are those where 
the low ionization ions are found, a density bounded model
has lower absolute intensities and a higher average ionization degree
than the corresponding radiation bounded model.
The second possibility to model photon escape in
a simple way is to assume that the 
cloud does not cover the whole solid angle around the 
ionizing source, but rather
an angle $\Omega < 4\pi$,
a situation expressed by stating that a {\it covering factor}
is assumed;
the covering factor is the fraction $cf = \Omega/4\pi$.
The physical picture is that all the ionizing photons 
emitted along certain directions are completely trapped,
whereas all those emitted in other directions freely escape;
or, equivalently, that the model 
is radiation bounded along certain directions, and 
`extremely' density bounded (in the sense that no gas at all is encountered)
along the others.
The model is still spherically symmetric because the 
two kinds of directions are assumed to be completely mixed.
A model with $cf < 1$ has
lower absolute intensities and the same ionization degree
as the corresponding model with $cf = 1$,
because low and high ionization degree zones are cut away proportionally.
Real nebulae are probably intermediate between the two scenarios
described: 
the gas ends up at different radii according to the direction
considered, possibly covering all the range between 0 and $R_S$.

Note that the simple relations between the properties
of radiation and density bounded models, 
and between the properties of models
with $cf = 1$ and $cf < 1$, hold only if 
all the other input parameters are left unchanged.
In the actual modeling practice things are not so straightforward,
since we might want to change simultaneously several input parameters
depending on which constraints we adopt. For instance,
if we want to reproduce a given total $I({\rm H}\beta)$ value
to fit an observed H{\sc~ii} region, 
and to this aim we compute 
a sequence of models with varying $cf$,
the observational constraint on  ${\rm H}\beta$ forces us
to increase the ionizing flux proportionally to the decrease in $cf$,
and this change simultaneously yields changes in the local ionization degree.

\subsection{Output quantities\label{sec:output_quantities}}

For each model, we computed the intensities of the most important 
emission lines 
(H$\alpha$, H$\beta$, 
[O{\sc~iii}] $\lambda\lambda\,5007, 4363$,
[O{\sc~ii}] $\lambda\lambda\,3727, 7325$,
[O{\sc~i}] $\lambda\,6300$,
[S{\sc~ii}] $\lambda\lambda\,6716, 6731$,
and 
He{\sc~ii} $\lambda\,4686$);
the average temperatures $T_0({\rm H}^+)$, $T_0({\rm O}^+)$, and $T_0({\rm O}^{++})$
(hereinafter $T_0$, $T_{02}$, and $T_{03}$, as in Paper I),
with $T_0({\rm X})$ defined, for a given ion X, by the expression:
\begin{equation}
T_0({\rm X}) = \frac{\int_V T_{\rm e} N_{\rm e} N({\rm X}) dV}{\int_V N_{\rm e} N({\rm X}) dV},
\end{equation}
where $T_e$ is the electron temperature,
$N_e$ the electron density,
$N({\rm X})$ the ionic density, 
and $V$ the observed volume; 
the `empirical' [O{\sc~iii}] temperature $T(4363/5007)$
(hereinafter $T({\rm O~{\scriptstyle{III}}})$)
obtained from the standard relation applied to
the predicted $\lambda\,4363/\lambda\,5007$ ratio;
the `empirical' [O{\sc~ii}] temperature $T(3727/7325)$
(hereinafter $T({\rm O~{\scriptstyle{II}}})$),
obtained from the expression
\begin{equation}
T({\rm O~{\scriptstyle{II}}}) = T_{02}\left[ 1 + \left(\frac{97,300}{T_{02}}-3 \right) \frac{t_2^2}{2} \right]
\label{eq:TOII}
\end{equation}
\citep{P67};
the temperature fluctuation parameters $t^2({\rm H}^+)$, $t^2({\rm O}^+)$, and $t^2({\rm O}^{++})$
(hereinafter $t^2$, $t^2_2$ and $t^2_3$, as in Paper I),
with $t^2({\rm X})$ defined as
\begin{equation}
t^2({\rm X}) = \frac{\int_V [T_{\rm e} - T_0({\rm X})]^2 N_{\rm e} N({\rm X}) dV}{T_0({\rm X})^2\int_V N_{\rm e} N({\rm X}) dV};
\end{equation}
the ionization fraction ${\rm O}^+/{\rm O}$, 
defined as 
\begin{equation}
{\rm O}^+/{\rm O} = 
 \frac{\int_V N_{\rm e} N({\rm O}^+) dV}{\int_V N_{\rm e} [N({\rm O}^+) + N({\rm O}^{++})] dV},
\end{equation}
called $\alpha$ in Paper I; 
the related quantity\footnote{
Note that these quantities are {\it not} the same as those 
listed in the `Log10 Mean Ionisation (over volume)' block 
at the end of the output of Cloudy 94,
which are simple volume averages of the ionization fractions;
in the case of non-constant density the two definitions do not
coincide. For an outward-decreasing density law, Cloudy's definition
yields lower fractions for the high ionization degree ions.}
${\rm O}^{++}/{\rm O} = 1 - {\rm O}^+/{\rm O}$;
and the helium ionization correction factor, $icf({\rm He})$,
defined by the relation:

\begin{eqnarray}
\frac{N ({\rm He})}{N ({\rm H})} & = &
\frac {\int{N_e N({\rm He}^0) dV} + \int{N_e N({\rm He}^+) dV} + 
\int{N_e N({\rm He}^{++})dV}}
{\int{N_e N({\rm H}^0) dV} + \int{N_e N({\rm H}^+) dV}},
                                                \nonumber \\
& = & icf({\rm He})
\frac {\int{N_e N({\rm He}^+) dV} + \int{N_e N({\rm He}^{++}) dV}}
{\int{N_e N({\rm H}^+) dV}}
\label{eq:ICF}
.\end{eqnarray}

The emission spectrum and the other computed quantities 
of each model were modified to simulate
the bias introduced by the angular extension of the slits 
used in the observations.
This step is crucial in the modeling,
since the observed emission spectrum depends strongly on the 
portion of the nebula included in the aperture,
as a consequence of the ionization and temperature structure of the nebula:
see a thorough discussion and multiple examples
of this effect in \citet{LPL99} and \citet{LP01}.

The output quantities fall in three distinct categories:
the first group includes those quantities
that are computed to be compared with the corresponding observational constraints,
and their agreement determine whether the fitting procedure
is converging or not;
this is the case of the line intensities, 
and of $T({\rm O~{\scriptstyle{II}}})$ and $T({\rm O~{\scriptstyle{III}}})$,
which are the theoretical quantities that correspond to the observed
temperatures derived from the line ratios 
[O{\sc~ii}] $\lambda\,3727/\lambda\,7325$ and [O{\sc~iii}] $\lambda\,4363/\lambda\,5007$.
To this respect, note that the intensities of the helium lines
have not been included in the set of observational constraints
because they are not useful for this scope:
as all the recombination lines arising from a dominant ion,
they are very weakly dependent on the features of the model.
The second group includes those
output quantities computed with the aim of obtaining
an information that is not available from observations:
this is the case of $icf$(He),
whose predicted value is assumed to be equal to the actual one,
and is used to compute the total helium abundance.
Finally, the remaining quantities are computed 
either for information, or because they are necessary
to compute other predicted quantities:
for example, 
$T_{02}$ and $t^2_2$ are used to compute the predicted 
$T({\rm O~{\scriptstyle{II}}})$ by means of equation~(\ref{eq:TOII}).
The average temperatures $T_{0}$, $T_{02}$, and $T_{03}$
and the temperature fluctuations parameters 
$t^2$, $t^2_2$, and $t^2_3$  give a measure of the 
temperature structure of the models,
and describe how it compares with the temperature structure of the actual region. 
As a rule, $t^2$, $t^2_2$, and $t^2_3$ always turn out to be
smaller than the corresponding empirical quantities.
Analogously, $T_{02}$ and $T_{03}$ are
always smaller than $T({\rm O~{\scriptstyle{II}}})$ and $T({\rm O~{\scriptstyle{III}}})$
respectively,
unless $t^2_2=0$ or $t^2_3=0$;
see \citet{P67} and Paper~I for more details.

\subsection{Observational constraints\label{sec:obs_constraints}}

The basic conceptual step in the fitting procedure
is the comparison between the observed line intensities
to the value predicted by our models.
In this comparison, the largest weight is given
to H$\beta$, [O{\sc~iii}]~$\lambda\,5007$,
and [O{\sc~ii}] $\lambda\,3727$,
for the following two reasons:
first, these lines are among the brightest of the spectrum, 
so they carry the smallest errors;
second, they define the energetics and the ionization structure
of the region, giving a first-order description
of its overall structure.
On the other hand, 
[O{\sc~iii}] $\lambda\,4363$ is not a robust constraint,
because it may be affected by heating processes
other than photoionization \citep[see, e.g.,][]{SS99};
the intensity of [O{\sc~i}] $\lambda\,6300$ depends strongly
on the detailed description of the outer parts of the nebula,
and is enhanced by the occurrence of filaments and clumps;
finally, the detailed modeling of He{\sc~ii} $\lambda\,4686$ is not possible yet.
This line forms in the recombination of He$^{++}$,
an ion that can be produced potentially by different sources of radiation,
and each candidate source poses a different problem that
makes difficult to estimate quantitatively  
its relation to the nebular $\lambda\,4686$ line.
The most plausible sources of He$^{++}$ are the hard radiation field of
WR stars and the X-ray emission produced in the burst.
WR stars, which are revealed by broad emission features in the spectrum,
are not univocally associated to to nebular $\lambda\,4686$ emission,
since in some cases they are observed with no associated
nebular $\lambda\,4686$ line, and in other cases
the nebular $\lambda\,4686$ is observed without detected WR stars;
furthermore, the WR formation mechanism at low metallicities 
is still poorly understood, and 
standard evolutionary models systematically underpredict their number
\citep[see, e.g., the case of \object{I~Zw~18}:][]{Bal02}.
The second candidate, X-ray emission, can be 
in turn ascribed to different sources:
massive binary systems,
supernova (SN) remnants,
and the reprocessing of the gas kinetic energy
in the interstellar medium.
As for the first of these sources,
the expected number of massive X-ray binaries is roughly
4 percent of the number of newly formed O stars \citep*{DJF92};
a similar fraction is also obtained by means of
theoretical estimations 
\citep[M. Cervi\~no, M. Mas-Hesse, \& D. Kunth, in preparation; see also][]{MHal96}.
The presence of massive X-ray binaries also implies the existence
of SN remnants, which are a further source of He$^{++}$;
for both these sources to be present, the age of the burst must be larger than 3 Myr.  
Finally, the conversion of kinetic energy in X rays can be modeled 
assuming an efficiency factor, which can be estimated
from the size of the bubbles in the region and from
quantitative X-ray observations. Unfortunately,
these data are not available for the H{\sc~ii} regions considered
in this work,
with the only exception of \object{I~Zw~18}, where
X rays have been detected \citep{SS98}.

This brief discussion illustrates the difficulty of using
the nebular $\lambda\,4686$ intensity to constrain
the properties of the cluster. 
This difficulty is increased by the statistical dispersion in
both the number of WR stars and of X-ray
sources \citep*{CMHK02}:
this implies that, even if the quantitative relation between 
a particular source and the intensity of $\lambda\,4686$ were perfectly known, 
there would still be a residual uncertainty in the modeling of  
those regions where sampling effects are relevant.

\subsection{Collisional effects\label{sec:collisional_effects}}

The effects of the collisional enhancement of 
Balmer lines on the determination of $Y_P$
have already been discussed by several authors:
see, for example, \citet{DK85}, \citet{SK93}, \citet{SI01}, and \citet{PPL02}.
This section will give a general overview of the problem,
and describe the particular strategy followed in this work
to give a quantitative estimate of the effect.

\subsubsection{Theoretical overview\label{sec:collisions_theorovw}}

In H{\sc~ii} regions, the main process
contributing to the intensity of Balmer lines is
H$^+$ recombination. 
Under appropriate physical conditions, 
a further contribution can arise from
the cascade following the collisional excitation of H$^0$
from the ground state to an excited level.
This contribution is generally a minor one,
both because the fraction of H$^0$ inside the H{\sc~ii} region
is very small,
and because very few free electrons have
energies sufficient to make ground-state H$^0$ electrons jump to 
excited levels in the typical conditions
of H{\sc~ii} regions (a minimum of 12 eV is required to produce
H$\alpha$, more than that for the higher Balmer lines).
Nevertheless, collisional enhancement of Balmer lines
might reach a relative value of several percent.
If not taken into account,
the collisional contribution has two biasing effects 
on the determination of $Y_P$.
The first is a straightforward consequence of
the enhancement of the observed Balmer intensities:
if the observed intensity is interpreted in terms 
of pure H$^+$ recombination,
the He$^+$/H$^+$ abundance is underestimated.
The second effect is a consequence of the different collisional enhancements
suffered by H$\alpha$ and H$\beta$: since the relative collisional
increase in H$\alpha$ is always larger than that in H$\beta$, 
the measured H$\alpha$/H$\beta$
ratio is larger than it would be without collisions,
producing a spurious reddening in the Balmer spectrum
that mimics a higher extinction \citep{FO85}.
This extra-reddening, when not properly subtracted out
from the observed reddening to get the true extinction,
has different consequences on the abundance determination
according to the wavelength considered:
it enhances the ratio $I(\lambda)/I({\rm H}\beta)$
for lines blueward of H$\beta$, 
and decreases it for lines redward of H$\beta$.
Since He$^+$/H$^+$ is generally obtained as the weighted mean of 
the helium abundances obtained by different line ratios,
the net result of neglecting the effects of collisional enhancement 
depends on which helium lines are used;
in practice, it always traduces in a spurious $Y$ decrease
since the helium lines redward of H$\beta$ are globally
brighter than those blueward of H$\beta$
(particularly He{\sc~i} $\lambda\,5876$),
and therefore are those that weigh the most in the analysis.
An accurate determination of $Y$ relies therefore
on an accurate determination of the collisional effects
affecting the line intensities of the region considered.
This is in general a very difficult task,
as we will make clear in the following.

The ratio between the collisional and the recombination
contribution to the 
emissivity of a Balmer line, e.g. H$\alpha$, is 
related to the local conditions in the gas through the
following expression \citep[e.g.,][]{O89}:
\begin{equation}
\frac{j({\rm H}\alpha)_{\rm col}}{j({\rm H}\alpha)_{\rm rec}}\propto
\frac{N({\rm H}^0)}{N({\rm H}^+)}
\frac{e^{-\Delta E/KT_e}}{T_e^{1/2}\alpha({\rm H}\alpha)}\label{eq:CtoR},
\end{equation}
where $\Delta E$ is the excitation energy of the upper ${\rm H}\alpha$ level
($n=3$) from the ground level ($n=1$), and
$\alpha({\rm H}\alpha)$ is the recombination coefficient.
In this expression,
we omitted those terms that are either constant or slowly varying with $T_e$.
We also omitted the contribution to $j({\rm H}\alpha)_{\rm col}$
from electrons excited to levels $n>3$ that cascade
down to $n=3$ and finally produce an H$\alpha$ photon. 
Although simplified, this expression serves the scope of illustrating
the difficulty of obtaining an accurate determination
of collisional enhancement.

First, the ratio $j({\rm H}\alpha)_{\rm col}/j({\rm H}\alpha)_{\rm rec}$
is directly proportional to $N({\rm H}^0)/N({\rm H}^+)$,
which is poorly known because it cannot be directly measured.
The only possibility to estimate it is through a photoionization
model; a very good model is needed to this
scope, and it is important to assess to which extent
the results obtained are model-dependent.

Second, the ratio $j({\rm H}\alpha)_{\rm col}/j({\rm H}\alpha)_{\rm rec}$
depends strongly on the temperature, particularly through the Boltzmann
factor $exp({-\Delta E/KT_e})$. 
This dependence poses a difficult problem, 
since most photoionization models underestimate
the temperature observed in H{\sc~ii} regions
(e.g., \citealt{LPL99,SS99,LP01}; but see also \citealt{Oal00} for a different result).
This fact suggests that an extra-heating energy source
is operating in H{\sc~ii} regions. 
Several hypotheses have been proposed to explain the nature of
such energy source, but none of them has yet found
general agreement \citep[e.g.,][]{Bal91,DS00,SS01,LCB01}.
Most important, even the knowledge of the nature of the
the extra-heating source would not probably imply a knowledge of
the exact temperature structure, and particularly
its spatial correlation to the ionization structure, 
which enters equation~(\ref{eq:CtoR}) through
the product of $N({\rm H}^0)/N({\rm H}^+)$ with
the Boltzmann factor.

Finally, an important issue is the accuracy of the atomic parameters
involved in the calculations. For the collisional transitions
from the hydrogen ground level up to $n=5$, 
Cloudy 94 uses the recent calculations by \citet{Aal00} and \citet{Aal02}.
Although these authors do not give any explicit number regarding the 
estimated accuracy of their calculations, they perform a comparison
with previous results, finding differences smaller than 15 percent
in the energy range of interest. Whether this number can be taken
as a good measure of the current uncertainties, it is difficult to say;
our impression is that the overall uncertainty introduced by atomic
physics plays a minor role in this study.

\subsubsection{Observational constraints to collisional enhancement\label{sec:collisions_obsconstraints}}

An upper limit to the collisional enhancement of Balmer lines
is always implicitly defined by the observed reddening.
This limit corresponds to the extreme case in which
the internal extinction of the observed object,
$C({\rm H}\beta)^{\rm int}$, is negligible
(with ``internal extinction'' meaning extinction
caused by dust physically associated to the {\sc H~ii} region,
independently of it being mixed with the ionized gas or not),
and is therefore equal to the collisional enhancement
that would produce exactly the observed reddening,
corrected for the Galactic intrinsic extinction $C({\rm H}\beta)^{\rm gal}$:
that is, if we define a collisional reddening coefficient as
\begin{equation}
C({\rm H}\beta)^{\rm col}=\frac{{\rm Log}(I({\rm H}\alpha)^{\rm tot}/I({\rm H}\beta)^{\rm tot})
- {\rm Log}(I({\rm H}\alpha)^{\rm rec}/I({\rm H}\beta)^{\rm rec})}
{-f({\rm H}\alpha)},
\end{equation}
where $I(\lambda)^{\rm tot}$ and $I(\lambda)^{\rm rec}$ are the total 
and the recombination intensities, and
$f({\rm H}\alpha)$ is the value of the reddening function at ${\rm H}\alpha$,
then the observed reddening coefficient can be written as:
\begin{equation}
C({\rm H}\beta)^{\rm obs}=C({\rm H}\beta)^{\rm gal} + C({\rm H}\beta)^{\rm int} + C({\rm H}\beta)^{\rm col},
\end{equation}
thus
\begin{equation}
C({\rm H}\beta)^{\rm col} \le C({\rm H}\beta)^{\rm obs}-C({\rm H}\beta)^{\rm gal}.
\end{equation}
In the following, by the expression ``collisional reddening'' we will
always implicitly refer to the collisional reddening {\it of the Balmer lines},
and not to a global reddening affecting the whole spectrum.

It should be noted that the same line of reasoning
exposed above could be applied to higher Balmer lines, 
with the aim of obtaining further
constraints on collisional excitation.
The amount of estimated collisions has a weaker
direct influence on the derived H$\delta$ and H$\gamma$ 
intensities than in the case of H$\alpha$ and H$\beta$,
since these lines have higher excitation thresholds.
On the other hand, the degree in which collisions occur determines 
how much of the observed reddening is 
due to interstellar extinction, which in turn has 
an important impact on these lines:
therefore, they can be in principle strong indicators 
of collisions through the indirect effect on the
estimated extinction.
However, they have also larger relative observational uncertainties
as compared to H$\alpha$ and H$\beta$, so their actual power
to put constraints on collisions depends crucially
on the data quality. 
Since the whole procedure of estimating collisional contribution 
admittedly skates on thin ice, we chose in this work to rely solely on 
the more robust constraints provided by H$\alpha$ and H$\beta$.

\subsubsection{The approach of photoionization modeling\label{sec:collisions_photoapproach}}

In this work, our approach to the quantitative estimation 
of collisional enhancement is the following: 
first, for each object we compute a best-fit model that reproduces
the observational constraints, paying particular attention
to the ionization structure;  
second, we compute, for that particular model,
the amount of collisional contribution to H$\alpha$ and H$\beta$;
third, we assess in each case the uncertainty affecting
our estimate of the collisional contribution. 
The last is by far the most difficult step, 
and there is no unique and completely safe way
to undertake it.
Our strategy to face this problem is to determine 
the range of collisional contribution spanned by 
satisfactory models of a given region (i.e., 
how much model-dependent is the collisional contribution),
and the range of collisional contribution spanned by
temperature-enhanced models.
The temperature-enhanced models necessary to
realize this last point have been computed with different
modeling approximations.
The outcome will be discussed in detail for the three cases
in the next sections.

\section{\object{SBS~0335--052}\label{sec:SBS0335}}

\object{SBS~0335--052} is the second most metal-poor galaxy known 
with $Z = 1/40$ Z$_\odot$, 
a metallicity only slightly higher than the one of \object{I Zw 18}.
We will start our discussion with this object,
because the observational constraints available
for \object{SBS~0335--052} are by far more numerous and articulate
than for the other two objects. 
This fact allowed us to build more sophisticated models,
and to determine in a more stringent way the collisional contribution
to the Balmer lines and the corresponding uncertainties.

\subsection{Observational constraints}\label{sec:SBSobs_const}

\object{SBS~0335--052} is a roundish region with a redshift $z = 0.0136$ 
\citep{Ial97b},
yielding a distance $d=57$ Mpc for $H_0$=72 km s$^{-1}$ Mpc$^{-1}$.
\citet*{ICS01} report observations of ionized gas extending
over a region of $\sim 26'' \times 32''$,
corresponding to linear dimensions $\sim 7200\;{\rm pc} \times 8800\;{\rm pc}$,
but the H$\beta$ intensity falls to less than 1/100 of its peak value
within a region of angular radius $\sim 3.5''$ \citep[Figure 2a in][]{ICS01},
corresponding to a linear radius of $\sim 1000$ pc,
so we will adopt this value as an approximate constrain on the radius.

The total H$\beta$ flux\footnote{As customary, we will indicate with 
$F(\lambda)$ the observed fluxes, 
and with $I(\lambda)$ the dereddened (intrinsic) fluxes, 
both in units erg s$^{-1}$ cm$^{-2}$.} observed by \citet{Ial99}
is $F({\rm H}\beta) = 5.93 \times 10^{-14}$ erg s$^{-1}$ cm$^{-2}$;
the actual flux is certainly higher, 
because the slit they used 
(a $1'' \times 5.4''$ extraction of a longer slit: see below) 
does not cover the whole nebula.
The reddening coefficient $C({\rm H}\beta)^{\rm obs}$ determined by \citet{Ial99}
varies from 0.225 dex to 0.33 dex along the slit; 
we will adopt in the following the value $C({\rm H}\beta)^{\rm obs}=0.26$ dex,
corresponding to the total $F({\rm H}\alpha)/F({\rm H}\beta)$ in the slit.
Adding up the fluxes observed by \citet{Ial97b} and \citet{Ial99},
and subtracting out the portion common to the slits used in the two works
(which were perpendicular to one another),
we derive as a lower limit to the total intrinsic
flux emitted by the
region $I({\rm H}\beta)\sim 2.16 \times 10^{-13}$  erg s$^{-1}$ cm$^{-2}$,
or a luminosity $L({\rm H}\beta)\sim 8.4 \times 10^{40}$  erg s$^{-1}$.
This luminosity corresponds to $Q({\rm H}^0) = 1.8 \times 10^{53}$ s$^{-1}$
ionizing photons; 
this number also is a lower limit
to the actual number of ionizing photons emitted by the central source,
both because it is derived from a lower limit on $I({\rm H}\beta)$, 
and because photons may leak out of the region without producing any
${\rm H}\beta$.
Adopting a conservative value of 20 percent photon leak,
we estimate $Q({\rm H}^0) \grsim 2.2 \times 10^{53}$ s$^{-1}$.
The root mean square (rms) density, derived from the observed H$\beta$ flux, is 
$N_{\rm e}({\rm rms}) = 4.8 {\rm~cm}^{-3}$;
for a filling factor  $\epsilon \sim 0.01$, which is a typical value
for giant H{\sc~ii} regions, this would correspond
to an average local (forbidden-line) density $N_{\rm e} \sim 48 {\rm~cm}^{-3}$.
This agrees with the data reported by \citet{Ial99}, 
who derived, from the [S{\sc~ii}]~6717/6731 ratio,
density values ranging from 500 ${\rm cm}^{-3}$ near the center
to 10 ${\rm cm}^{-3}$ in the outer zone.

The nebular line He{\sc~ii} $\lambda\,4686$ is observed
in \object{SBS~0335--052}, with an intensity relative to H$\beta$
in the range (0.02 -- 0.05) depending on the position,
with the highest values observed in the brightest part of the nebula.
As possible sources for He{\sc~ii} $\lambda\,4686$,
\citet{Ial97b} consider WR stars, MS stars, and X-ray binaries. 
Since they do not observe any broad WR feature in the spectrum,
and the hard emission from MS stars does not seem 
intense enough to produce the observed $\lambda\,4686$ flux,
these authors favor the hypothesis of X-ray binaries
(however, they do not consider the X rays produced
by stellar winds and SN explosions through
the release of kinetic energy,
neither the X-ray emission of SN remnants,
although the presence of X-ray binaries implies,
at least, one previous SN explosion: see the discussion in Section~\ref{sec:obs_constraints}).
An additional argument invoked to rule out WRs is the spatial distribution 
of $\lambda\,4686$ across the nebula, 
which is displaced from the other nebular lines;
indeed, \citet{ICS01} find a spatial correlation
of the line intensity with the position of a supershell
that is probably a shocked region.
a more general argument is that WRs formed
through single-star evolution are expected to be very rare
at extremely low metallicities.
Nevertheless, the spectra taken by \citet{Ial99}, 
at a position angle perpendicular to the previous one,
show a weak bump around 4620 -- 4640 \AA,
suggesting the presence of WR stars, probably of the WN type.
In this work, we do not make any attempt to reproduce
the nebular $\lambda\,4686$, due to the loose connection between 
the WR stars and nebular $\lambda\,4686$ emission
that has been discussed in Section~\ref{sec:obs_constraints}.
On the other hand, we take advantage of the observed stellar 
WR features to constrain the age of the region;
if the WRs of \object{SBS~0335--052}
have formed through the standard, single-evolution
channel, the region must be about $3$ Myr old.

The observational constraints used in our modeling
are the 
line ratios obtained by \citet{Ial99} 
with the Low-Resolution Imaging Spectrometer of the Keck II telescope.
They used a $1''\times 180''$ slit, 
centered on the second brightest cluster of the region
and oriented SW-NE.
{F}rom this slit, nine extractions of size $1''\times 0.6''$ were obtained.
Our aim is to reproduce the line ratios
in the nine extractions simultaneously;
this corresponds to a formidable number of independent, but interrelated,
observational constraints, which altogether provide a much
firmer handle on the inner structure of the nebula than the integrated properties.

\subsection{The models of \object{SBS~0335--052}}

\subsubsection{General features}

Our models of \object{SBS~0335--052}
are spherically symmetric and have a
Str\"omgren radius of approximately $R_S \sim 1000$ pc.
We assume $Z_* = 0.001$ and $Z_{gas} = 0.0005$
for the stellar and gas metallicity respectively.
Since the region is very large, and the stars appear
scattered all over the projected surface (although
they are preferentially concentrated towards the center),
we assume that the star formation is going on in a continuous
burst, which began $t = 3.0$ Myr ago.
To compute the ionizing spectrum, we adopted the following procedure:
we performed 100 Monte Carlo simulations with 2$\times 10^4$ stars each,
assuming an IB case with a time step of 0.1 Myr;
each cluster was evolved up to 3.0 Myr.
The resulting mean mass of these synthetic clusters 
is 1.18 $\times 10^5$ M$_\odot$ in the range 2 -- 120 M$_\odot$.
In order to obtain the spectrum of a CSF burst, these simulations 
have been summed over time taking randomly a Monte Carlo
simulation for each age.
The mean total stellar mass of the synthetic clusters is 3.5 $\times 10^6$ M$_\odot$, 
which gives a CSF rate of 1.18 M$_\odot$ yr$^{-1}$. 
Due to the large total mass of this region,
the statistical fluctuations in the sampling of the IMF are very small,
and the resulting ionizing spectrum obtained with this procedure is almost
identical to the one computed assuming a completely sampled IMF.

We adopt as a lower limit to the number of ionizing photons 
$Q({\rm H}^0) = 2.2 \times 10^{53}$ s$^{-1}$,
a choice leaving us the possibility to increase it
to improve the fit of any given model (see Section~\ref{sec:SBSobs_const});
this does not pose any theoretical problem
from the point of view of the IMF sampling, because
the IMF is well sampled.
In the gas we assume scaled-down solar abundances
for all the heavy elements, with the exception of nitrogen and carbon:
for these elements, the adopted abundances relative to solar are
1/160 and 1/120 respectively.
The helium abundance is $N({\rm He})/N({\rm H}) = 0.076$.

\subsubsection{The Gaussian model}\label{sec:SBS:gaussian}

The starting point for our computations was a radiation bounded
model with constant density $N_e = 50$ cm$^{-3}$,
filling factor $\epsilon = 0.01$, and
covering factor $cf = 1.00$.
Based on this reference model, we computed sequences of models
varying several parameters: $Z_{\rm gas}$,
$\epsilon$, $cf$, and the density law;
both constant and non-constant density laws were considered,
with the last case represented by a Gaussian law:
$N_{\rm e} = {\rm max}\,(N_{\rm e}^{\rm max} {\rm exp}(r/r_0)^{-2}, N_{\rm e}^{\rm min})$.

\begin{figure}
\plotone{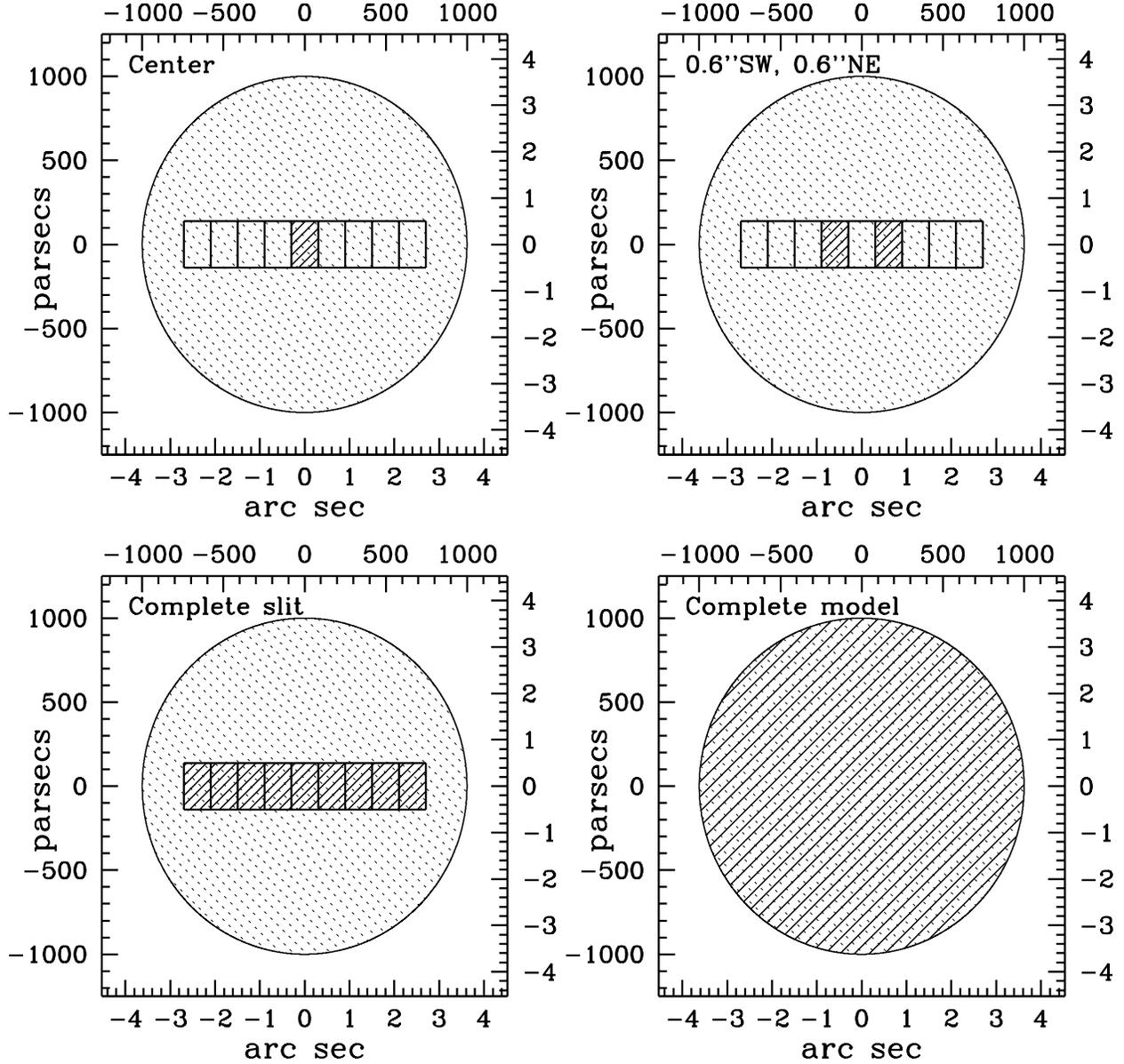}
\caption{Sketch of the slit used by \citet{Ial99}
and the nine extractions making it up, projected
onto the model nebula. 
This figure also provides a visual explanation for the
meaning of the column headers of Tables~\ref{tab:SBScomparison}
and \ref{tab:SBSpredicted}. 
The left and lower axes of each panel are in arc sec,
the right and upper axes are in parsecs
and assume a distance $d=57$ Mpc (see Sections~\ref{sec:output_quantities}, 
\ref{sec:SBSobs_const}, and \ref{sec:SBSmultipleshell}).
\label{fig:slits}}
\end{figure}

The output of each model is modified to simulate observations
through the nine extractions of the slit used by \citet{Ial99}
(Figure~\ref{fig:slits}).
Most of the weight in the comparison is given to the three most central
extractions, since their line intensities have the smallest errors.
The best-fit model obtained from this grid of models,
called hereinafter `Gaussian model',
has the following characteristics: $Z_{gas} = 0.0007$;
Gaussian density law, with 
$N_{\rm e}^{\rm max} = 720$ cm$^{-3}$, $N_{\rm e}^{\rm min} = 90$ cm$^{-3}$,
and $r_0 = 375$ pc; $\epsilon = 6.4\times 10^{-4}$; $cf = 0.80$.
The remaining parameters are the same as in the reference model 
described above. 

\begin{figure}
\plotone{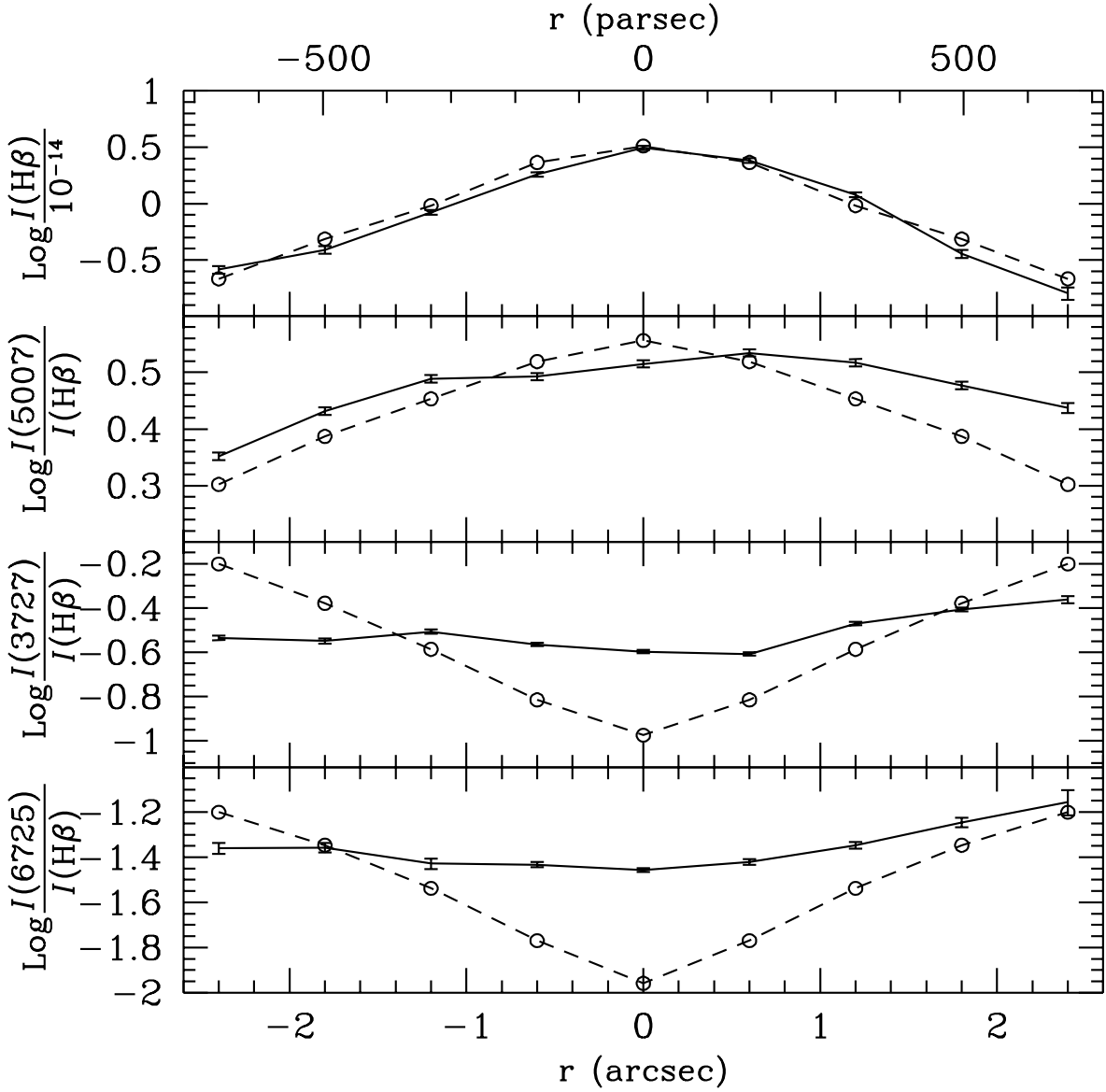}
\caption{Line profiles for the Gaussian model of \object{SBS~0335--052}.
Solid lines: observations, dashed line: model. 
The H$\beta$ intensity is in erg s$^{-1}$ cm$^{-2}$.
\label{fig:SBSgauss}}
\end{figure}

Figure~\ref{fig:SBSgauss} shows the comparison between the observed and the
predicted line profiles for this model; the points
correspond to the line intensities 
either observed or predicted in each of the nine extractions making up
the slit. 
The model is, by construction, symmetric with respect to the center,
whereas the observed profiles show mild asymmetries:
in the modeling, a fit is considered satisfactory whenever
the predicted point at any given radius
reproduces the average observational point at that radius,
obtained averaging the corresponding SW and NE values;
the observational uncertainties are shown as errorbars 
superposed on the observed profiles.
The top panel shows the H$\beta$ intensity, which is, 
within the observed uncertainties, very well reproduced by the model.
Nevertheless, the three bottom panels show large discrepancies
in the projected radial behavior of the ionization degree. 
These discrepancies are of special concern because
an accurate estimate of collisions requires reproducing
the ionization structure of the observed region.
The observed ionization degree is almost constant across the region,
a condition which is difficult to reproduce with a radiation bounded, spherical model
with a central source of radiation.
Physically, the observed profiles can be explained in several ways,
for example:
(i) the region is density bounded, so the low-ionization zone is cut off;
(ii) the local density in the outer zones is much lower than assumed by our model,
   implying a drop in the local recombination rate and,
   consequently, a higher ionization degree in the outer zones;
(iii) the ionizing stars are scattered throughout the region, each creating
   its own smaller H{\sc~ii} region with high- and low-ionization zones, 
   the complex of H{\sc~ii} regions merging with each other.

Alternative (i) can be excluded on the basis of the relatively low ionization degree,
since in a density bounded region the ionization degree is smooth but high. 
Alternative (ii) would imply a simultaneous increase in the filling factor
of the outer zones, to preserve the observed H$\beta$ flux;
but such possibility is not too appealing, since the scarce
observational evidence regarding this point seems to suggest
either the opposite behavior, or a nearly constant filling factor \citep[e.g.,][]{O89}.
Furthermore, the outer density of the Gaussian model is already quite low, 
so that no much room is left for lowering it further. 
Alternative (iii) is supported by a cursor inspection of direct images
of \object{SBS~0335--052} \citep*[see, e.g.,][]{TIL97},
and is also the most consistent with 
the scenario of a CSF burst. 
Direct images also show that the density structure
of the region is filamentary, rather than smooth.
To take these facts into account and obtain a better fit,
we built a more complicated model that simulates
the filamentary distribution of the gas inside \object{SBS~0335--052}, 
and the distribution of the ionizing sources across the cloud.
This model will be described in the next section.

\subsubsection{The multiple-shell model}\label{sec:SBSmultipleshell}

In this section we present a different model that
improves the fit notably. 
The new model is the sum of a sequence of thin, constant-density  
shells, with different radii,
and it will therefore be called `multiple-shell model'.
Since three alternative versions of this model will be presented,
all with the same geometrical layout,
we will further refer to them as A, B, and C;
multiple-shell model A is the reference one and is described in the following.

\begin{figure}
\plotone{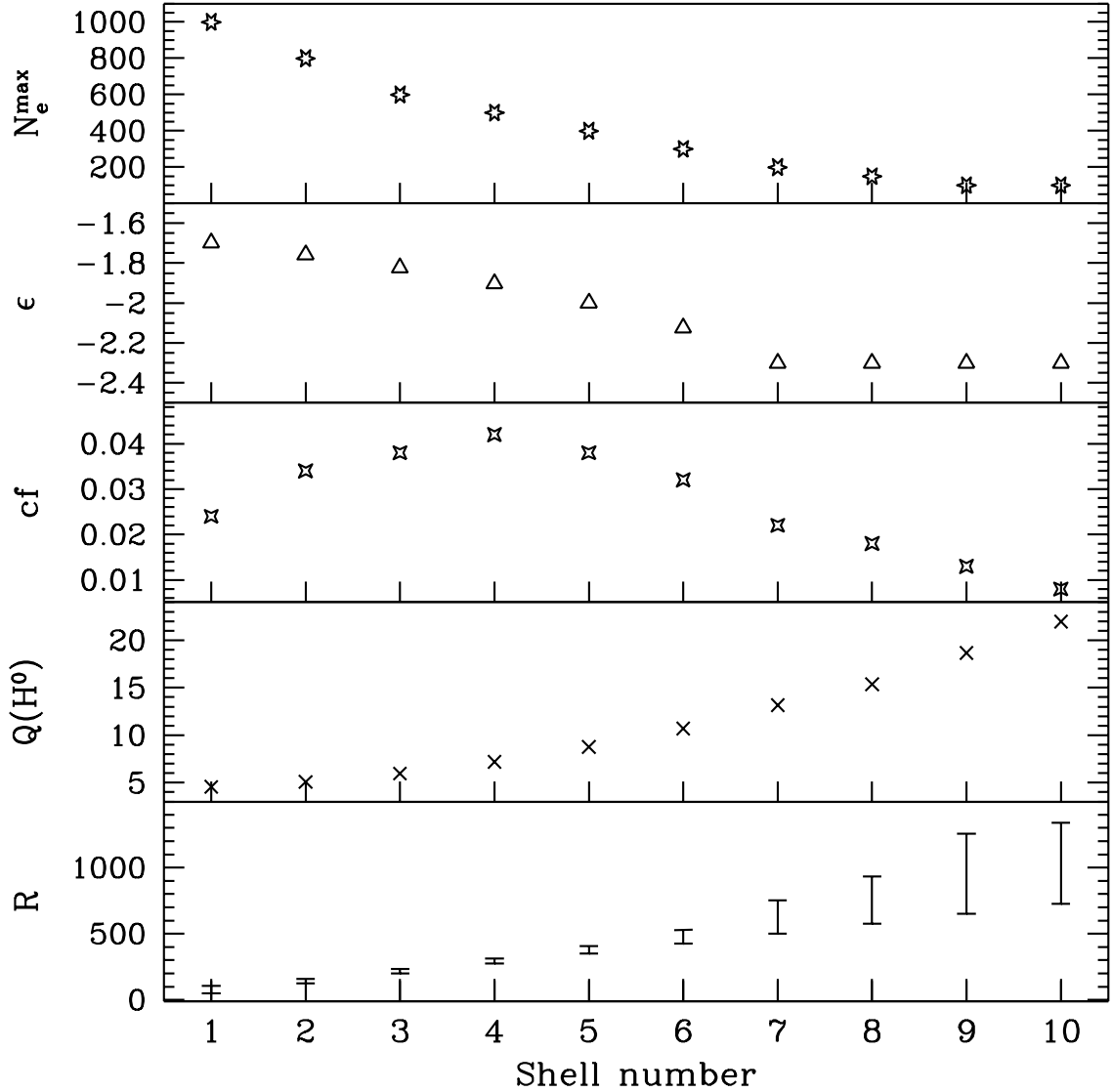}
\caption{Input parameters of the ten shells that compose
the multiple-shell model A.
The density law in each shell is given by
$N_{\rm e} = {\rm max}\,(N_{\rm e}^{\rm max} {\rm exp}(r/r_0)^{-2}, N_{\rm e}^{\rm min})$,
with $r_0 = 260$ pc, $N_{\rm e}^{\rm min} = 90$ cm$^{-3}$ equal in all the
the shells, and $N_{\rm e}^{\rm max}$ given by the values plotted in the upper panel.
The second panel from the top gives the filling factor values
of each shell;
the third panel from the top gives the covering factor values;
the effective covering factor, averaging $\langle cf \rangle=0.22$
over the model.
The fourth panel from the top gives the rate of ionizing photons
emitted, in units 10$^{53}$ s$^{-1}$; 
finally, the bottom panel shows the radius range spanned by each shell, in parsecs.
\label{fig:SBSparameters}}
\end{figure}

\begin{figure}
\plotone{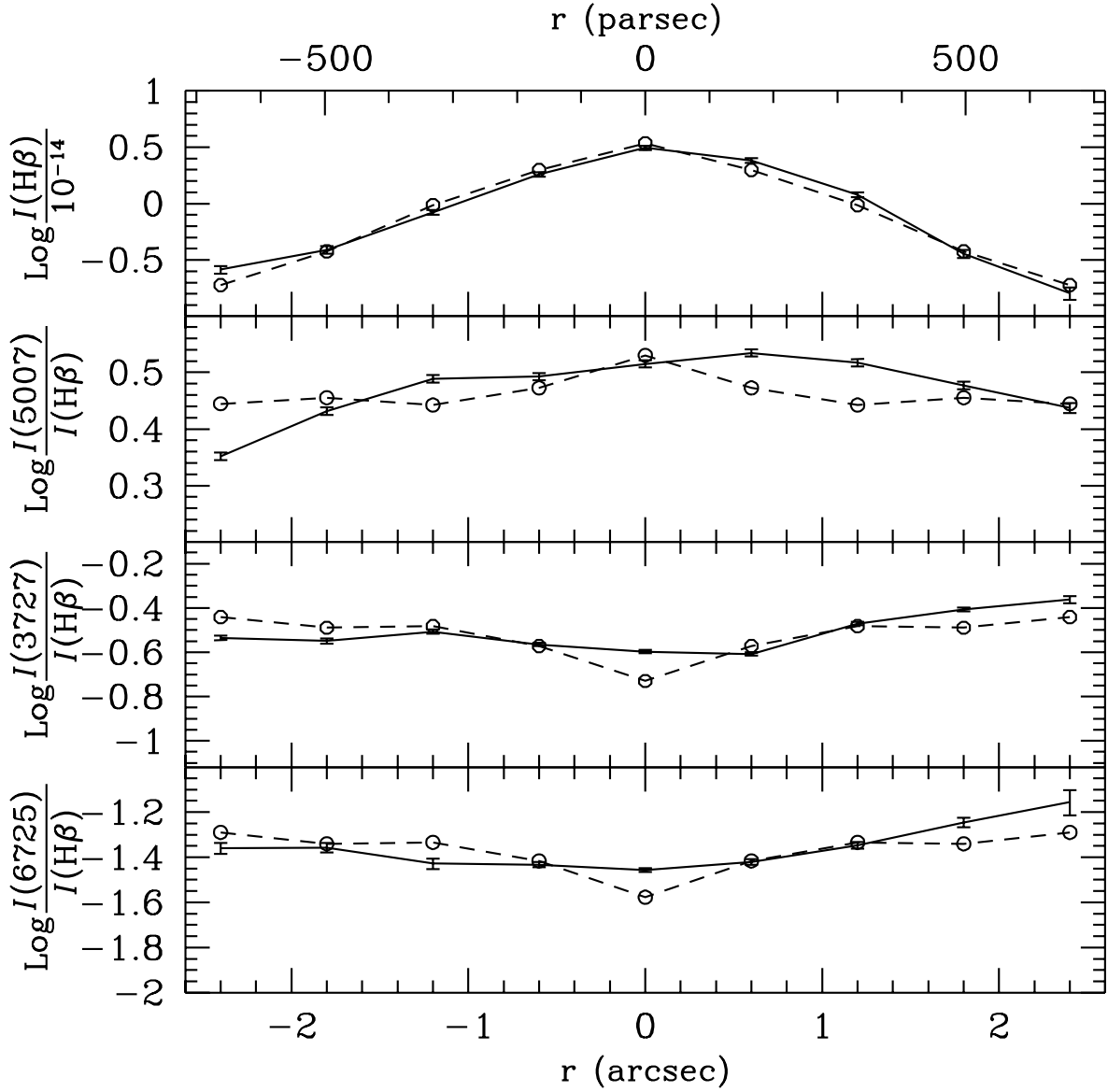}
\caption{Line profiles for the best multiple-shell model A of \object{SBS~0335--052}.
Solid lines: observations, dashed line: model.
The H$\beta$ intensity is in erg s$^{-1}$ cm$^{-2}$.
\label{fig:SBSshell}}
\end{figure}

Each shell composing model A is in itself radiation bounded, 
and has $cf < 1$.
In each shell we could, in principle, independently vary the gas distribution
(local density, filling factor, 
and covering factor); additionally, we could also vary
the number of shells and their inner radii.
We decided rather to fix {\sl a priori} the number of shells,
their inner radii, and the densities, 
pushed by several considerations:
first, we tried to keep things as simple as possible.
Second, the observational data 
and the modeling experience obtained with the Gaussian model
provide a good initial guess for the density structure.
Third, we want the shells to fill altogether the whole observed region,
with no substantial gaps or underlapping, so 
that, when observed globally, the structure appears fairly smooth:
this equals to a constraint on the outer radii, 
providing directly, with the further constraint on the density,
the filling factor value for each region; the resulting filling factor
values decrease smoothly in the outer direction, 
in agreement with some observational results (see Section~\ref{sec:SBS:gaussian}).
Fourth, in order to obtain a smooth structure 
without too many computational complications,
we fixed the number of shells to be ten.
The covering factor of each shell, which is left as a free variable, 
is a simple way to describe the filamentary structure of the region:
at each radius, photons encounter voids and regions filled with gas,
mixed up together (so that, at the macroscopic level of the slit,
voids and clumps are intercepted proportionally to the volume
they occupy globally).
Finally, we still have to take into account that the ionizing stars
are scattered throughout the nebula. 
This is not possible to compute directly with a 1D code as Cloudy
(see \citealt{V03} for a discussion on the differences between 1D and 3D codes),
so we resorted to the following approximation: 
as far as we proceed
from the innermost to the outermost shell, we gradually increase 
the absolute luminosity of the ionizing source. 
This increase reflects the fact that, when stars are scattered throughout
the nebula, the outer zones receive
a higher flux relatively to the central zones
than it occurs in the case of a central source.
The absolute normalization is then achieved by tuning the 
covering factor of each shell.
The resulting effective covering factor, obtained as the luminosity-weighted average of
the covering factors of the individual shells, is $\langle cf \rangle=0.22$.
This shell structure gracefully reproduces the constancy of
the ionization degree along the diameter of the nebula: this is a consequence
of each shell having its own low- and high-ionization zones.
At the same time, the H$\beta$ flux profile is equally well
reproduced as in the Gaussian case, because the overall density
structure is nearly the same.
Figure~\ref{fig:SBSparameters} summarizes the input parameters
of model A, 
and Figure~\ref{fig:SBSshell} illustrates the line profiles obtained
when the model is observed through the slit.

\begin{figure}
\plotone{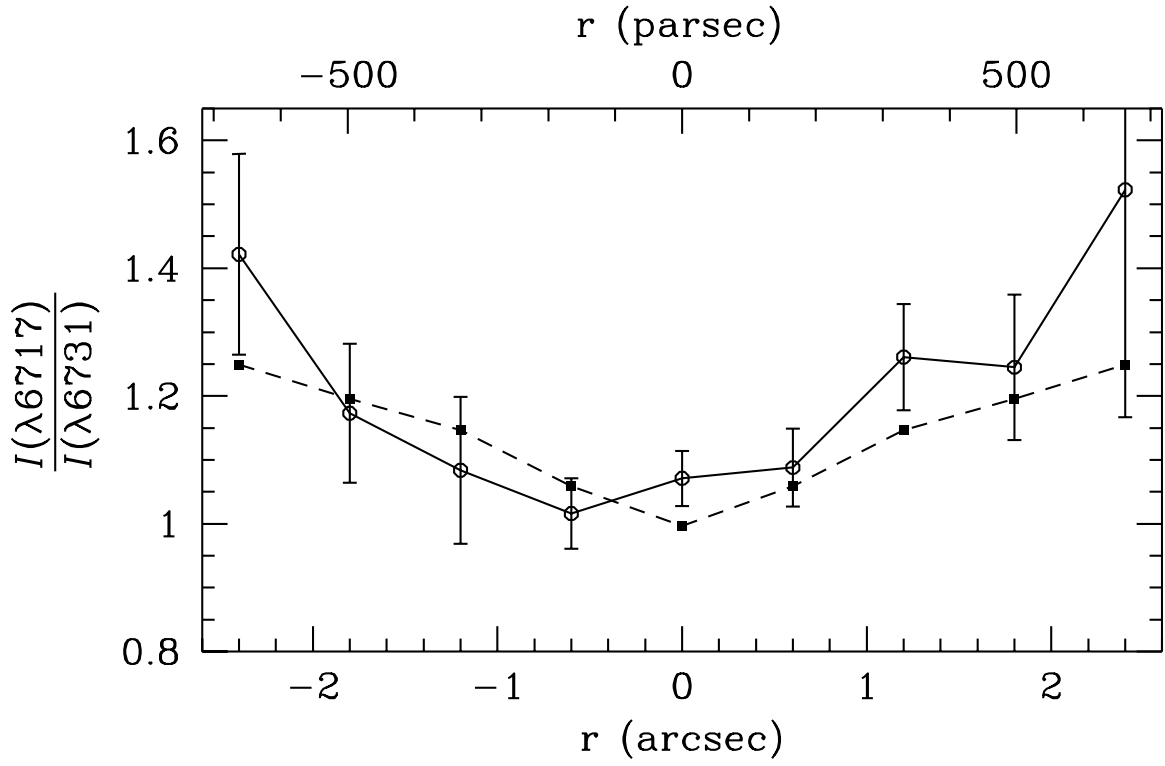}
\caption{Comparison between the observed [{\sc S~ii}] 6716/6731 ratio
profile (solid line) and the corresponding predicted values (dashed line)
for the multiple-shell model A of \object{SBS~0335--052}.
\label{fig:SBSratioSII}}
\end{figure}

Table~\ref{tab:SBScomparison} compares the predicted spectrum to the observational constraints.
The column header ``Center'' refers to the centermost extraction.
The label ``$0''.6$'' indicates, in the case of `Observed',
the average intensities of the two neighbouring extractions
\citep[indicated as ``$0''.6$SW'' and ``$0''.6$NE'' in Table~2 of][]{Ial99};
in the case of `Predicted', it indicates the intensities
predicted for either of the two extractions situated on both sides
of the central extraction.
``Complete slit'' indicates the sum of the nine extractions.
``Complete model'' is given for reference, 
and it indicates the results for the complete
model, with no aperture correction applied: 
no `Observed' column exists in this case,
since there are no observational data sampling the whole nebula.
The meaning of the column headers is also visually explained by Figure~\ref{fig:slits}.
The observed $T({\rm O~{\scriptstyle{III}}})$ values have
been computed from the line intensities of \citet{Ial99};
we did not use the temperature values reported by these authors,
since they only list them for single extractions,
i.e. they do not give a value for the complete slit.
For the same reason, the $T({\rm O~{\scriptstyle{II}}})$ values of
Table~\ref{tab:SBScomparison} differ slightly from the corresponding ones
given by \citet{Ial99}, since we recalculated them
in order to fill homogeneously all the slots of Table~\ref{tab:SBScomparison}.
In our calculation we assumed the same relation 
as \citet{Ial99} between  $T({\rm O~{\scriptstyle{II}}})$ and
$T({\rm O~{\scriptstyle{III}}})$,
which is based on a fit to the photoionization models by \citet{S90}.
Finally, Table~\ref{tab:SBSpredicted} contains 
various predicted quantities for model A,
and Figure~\ref{fig:SBSratioSII} compares the observed [{\sc S~ii}] 6716/6731
profile to the corresponding predicted values.

\subsection{Collisional effects in \object{SBS~0335--052}}\label{sec:SBScollisions}

\subsubsection{Model A}\label{sec:SBSmodelA}

Table~\ref{tab:SBScomparison} shows that the model systematically underpredicts 
the temperature of the region. 
This disagreement has been a systematic feature
in previous attempts of modeling H{\sc~ii} regions 
(see Section~\ref{sec:collisional_effects}),
and can be ascribed to an additional energy source
acting in photoionization regions,
other than photoionization itself.
It can be ignored in most cases when modeling
an H{\sc~ii} region,
provided the relevant line ratios are reproduced;
but for our aim,
differences of $\sim$ 3,000 $K$ between the observed 
and the predicted temperature cannot be neglected, 
because collisional rates are so sensitive to the temperature.
It is therefore necessary to assess how the difference
in temperature between the model and the real region
affects the collisional enhancement of hydrogen lines:
this problem is the topic of next section.

\subsubsection{Models B and C}\label{sec:SBSmodelsBC}

Starting from the multiple-shell model A, we 
follow two different strategies to estimate 
how the enhanced temperature would affect collisions.
The first strategy consists in computing a new multiple-shell model,
named `model B', 
identical to model A in all respects, with the only exception
of the metal abundances, which have been reduced by 1/100;
in this way, we suppress almost completely the cooling due to metals
and obtain higher temperatures.
An important feature of this model is that 
the line profiles are the same as in model A,
ensuring that the ionization structure
of model B is the same as in model A;
this, in turn, ensures that 
the dependency of the collisional rates on the ionization degree
is well reproduced, a necessary ingredient to draw 
any conclusion on the relevance of collisions.

The second estimate of the effect of temperature
on the collisional rates 
is a simple numerical experiment,
consisting in recomputing the collisional rates at
each point assuming a temperature 15 percent higher than the one of 
the original model.
It should be emphasized that
this is just a numerical trick and does not involve any 
change in the structure of the model: 
it does not even involve the computation of a new model, 
since its only effect is to output
collisional rates computed with enhanced temperature values;
the collisional rates used internally by the code to determine
the equilibrium conditions are left unchanged. 
This implies that the computed collisional rates are {\sl not}
consistent with the model structure. 
We named this model `multiple-shell model C'.

\subsubsection{Quantitative estimate of collisional enhancement in \object{SBS~0335--052}}

Table~\ref{tab:SBScollisions} compares the relative contribution
of collisions to the total H$\alpha$ and H$\beta$ intensity
for the three multiple-shell models.
Since model C is not self-consistent,
its extremely high collisional rates (roughly twice those
of model B, and four times those of model A)
are probably an artefact.
On the other hand, model B is totally self-consistent,
its [O{\sc~iii}] temperature is very close to the observed one,
and so is the observed ionization structure along the slit.
It seems reasonable therefore to assume that the collisional rates
found for this model are a good estimate of the real ones.
Since model B has almost no metals,
its temperature is the highest achievable
in a normal H{\sc~ii} region with photoionization alone.
This temperature is also the highest observed
in H{\sc~ii} regions
(e.g., \object{SBS~0335--052} is even hotter than \object{I~Zw~18}):
hence the collisional enhancement of this model possibly defines
an upper limit to what occurs in real H{\sc~ii} regions.
It is important to stress that there is a completely independent argument 
supporting the last statement:
since the collisional enhancement of Balmer lines mimics a higher extinction,
the observed reddening values define implicitly an upper limit
to the maximum amount of collisional enhancement of each object:
see Section~\ref{sec:collisions_obsconstraints}.

Considering the case of the complete slit,
we find from Table~\ref{tab:SBScollisions} that the collisional contribution to H$\beta$ 
might range from a minimum of 2.1 percent to a maximum of 3.5 percent
of the total intensity,
with the corresponding contribution to H$\alpha$
varying from 7.4 to 11.7 percent.
In the first case,
the observed $F($H$\alpha)/F($H$\beta)$ ratio should be corrected by a factor 0.946,
and in the second by a factor 0.915.
The average $C({\rm H}\beta)^{\rm obs}$ observed along the slit
should therefore be decreased from the observed value 0.26 dex to a collision-corrected
range (0.14 -- 0.19) dex to get the true extinction.
The collisional reddening coefficients computed for all the
cases considered are listed in the last row of Table~\ref{tab:SBScollisions}.

Reversing the reasoning, we could find a maximum collisional
enhancement for the H$\alpha$/H$\beta$ ratio by assuming
that the $C({\rm H}\beta)^{\rm int}$ of the region is negligible.
Correcting the $C({\rm H}\beta)^{\rm obs}$ value with 
the galactic extinction value by \citet*{SFD98} 
($E(B-V)=0.047$, or $C({\rm H}\beta)^{\rm gal}$ = 0.07 dex),
we find $C({\rm H}\beta)^{\rm col}_{\rm max}$ = 0.19 dex. 
This amount of collisional reddening could be obtained
with a 16 percent collisional enhancement in the ratio $I({\rm H}\alpha)/I({\rm H}\beta)$;
this in turn would put a rough upper limit on the collisional enhancement equal to
$I({\rm H}\beta)_{\rm col}^{\rm max}/I({\rm H}\beta)_{\rm tot} = 0.05$,
where it has been assumed, based on Table~\ref{tab:SBScollisions}, 
that $I({\rm H}\alpha)_{\rm col}/I({\rm H}\alpha)_{\rm tot} \sim 3.5 
\times I({\rm H}\beta)_{\rm col}/I({\rm H}\beta)_{\rm tot}$.

\section{\object{I~Zw~18}\label{sec:IZw18}}

\object{I Zw 18} is the most metal-poor galaxy known, with 
$Z = 1/50$ Z$_\odot$.
Given this extremely low metallicity value, it is a key object in
the primordial helium analysis, and it has been studied by many authors.
The redshift of the galaxy is $z=0.00254$ \citep*{ITL97},
which gives a distance $d = 10.6$ Mpc for
$H_0$=72 km s$^{-1}$ Mpc$^{-1}$; its apparent
size is roughly $5''\times 10''$, corresponding to
linear dimensions 250 pc $\times$ 500 pc at the assumed distance.
\object{I Zw 18} is made up of two main regions,
usually referred to as the south-east (SE)
and north-west (NW) knots.
The NW region is more extended and globally brighter than the SE region:
the H$\beta$ flux observed by \citet{Ial99}
is $F({\rm H}\beta)= 2.08 \times 10^{-14}$ erg s$^{-1}$ cm$^{-2}$
in the NW knot\footnote{The value published in the paper,
$F({\rm H}\beta)= 1.04 \times 10^{-14}$ erg s$^{-1}$ cm$^{-2}$,
is an erratum (Y. I. Izotov 2003, private communication)},
and $F({\rm H}\beta)= 1.49 \times 10^{-14}$ erg s$^{-1}$ cm$^{-2}$
in the SE knot;
the NW knot has been observed with a $4.2''\times 1.5''$ slit,
and the SE knot with a $3.5''\times 1.5''$ slit.
The nebular He{\sc~ii} line $\lambda\,4686$ is observed
in \object{I Zw 18}, with $I(\lambda\,4686)/I({\rm H}\beta) = 0.041$
in the NW region, and 0.009 in the SE region \citep{Ial99};
\citet{VIP98} observe $\lambda\,4686$ emission throughout the whole galaxy,
with intensities relative to H$\beta$ ranging from 0 to 0.080.
The spectra of the NW knot show evidence of WR stars
\citep{Ial97a,Lal97,Bal02},
whereas no stellar WR features have been observed to date in the SE knot.

In this work we model the emission from the SE knot;
although less conspicuous than the NW knot, the SE knot
is less affected by underlying absorption in the helium lines,
and is therefore more suitable for a precise determination of the 
helium abundance in the galaxy.
As a drawback of this choice, it turns out to be quite difficult 
to estimate the age. 
Our assumptions regarding this point will be discussed in the next section.

\subsection{Observational constraints}

\subsubsection{Line intensities and physical conditions}\label{sec:lines_IZw18}

We used as observational constraints 
the line intensity ratios by \citet{Ial99}.
As in the case of \object{SBS~0335--052},
we modified the model's output to take into account 
the $1.5'' \times 3.5''$ slit used
by \citet{Ial99}, which samples, at the assumed distance,
a region of 77 pc $\times$ 180 pc.

The total H$\beta$ flux observed by \citet{Ial99}
is a lower limit to the total H$\beta$ flux emitted
by the SE knot, since the SE knot has a diameter of 
approximately $4''$.
To estimate the total flux, 
we compared Figure~1 in \citet{Ial99} to
Figure~3 in \citet{Cal02}.
The comparison suggests that the slit used by 
\citet{Ial99} covered the regions 
labeled SE D1, SE D4, SE D5, SE D6, and SE D7 by \citet{Cal02}
(although no precise correspondence can be established,
due to the scale of the figures and the seeing, which was equal
to the width of the slit).
The sums of the corresponding H$\beta$ fluxes coincide to
better than 15 percent, providing a consistency check for this assumption.
Therefore, we will adopt for the total H$\beta$ flux emitted
by the region the one given by \citet{Cal02} for the 8 clumps
labeled ``SE D1'' through ``SE D8'',
which is 23 percent higher than the one measured by \citet{Ial99}.

Consistently with our assumption on the total observed H$\beta$ flux,
we computed the total $I({\rm H}\beta)$ as the sum
of the $F({\rm H}\beta)$ values observed by \citet{Cal02},
each corrected for the corresponding extinction value;
we obtain $I({\rm H}\beta) = 2.32 \times 10^{-14}$ erg s$^{-1}$ cm$^{-2}$,
a value 56 percent higher than the one by \citet{Ial99}.

\subsubsection{Stellar population}

The values of Log $Q({\rm H}^0)$ estimated from the
H$\beta$ flux, together with reasonable assumptions on the covering factor
($0.40 \lesssim cf \lesssim 1.0$),
range from 50.8 to 51.2; 
this would correspond to a total amount of gas transformed into stars of
$4.2\times10^3 \lesssim M / {\rm M}_\odot \lesssim 6\times10^4$
in the mass range 2-120 ${\rm M}_\odot$.
The lower limit corresponds to a 2 Myr burst with Log $Q({\rm H}^0)=50.8$,
or about $7\times 10^2$ stars,
and the upper limit to a 5 Myr burst with Log $Q({\rm H}^0)=51.2$,
or about 10$^4$ stars.
Because these values are quite small,
sampling effects in the IMF cannot be neglected.
The absence of WR stars in the cluster cannot
be used to bracket the age, 
since the models predict, in the most
favorable case, 1.6 WR stars on average 
and a probability of 20 percent of observing no WR stars;
thus, we have no way of knowing
whether no WR stars are observed because of the age
or because of stochastic effects.
Additionally, the spectra of the region show no evidence of 
Balmer jump \citep{Gal97}, ruling out the 
presence of old post-MS luminous stars.
Based on these pieces of evidence,
we have assumed that the main stellar population is 
composed of MS stars,
and have explored relatively young populations
($2.0\, {\rm Myr} \lesssim t \lesssim 3.7\, {\rm Myr}$).

\subsubsection{Reddening coefficient}\label{sec:IZw18CHb}

The reddening coefficient in \object{I~Zw~18} varies strongly
from knot NW to knot SE, and within the same knot.
For example, recently determined values for the NW knot range
from \chb\ = 0.010 \citep{SK93}, to 0.011 \citep{ITL97},
0.040 \citep{IT98}, and 0.115 \citep{Ial99},
while those for the SE knot range
from \chb\ = 0.20 \citep{SK93}, to 0.265 \citep{IT98}
and 0.015 \citep{Ial99}.
Recently, \citet{Cal02} observed \object{I~Zw~18}
with the HST/WFPC2 finding \chb\ values in the (0.00 -- 0.18)
range for the NW knot, and in the (0.00 -- 0.27) range for
the SE knot.
The average extinction coefficient for these data is
$C({\rm H}\beta) = {\rm Log} [I({\rm H}\beta)/F({\rm H}\beta)] = 0.10$ dex.

\subsection{The models of \object{I~Zw~18}}

\begin{figure}
\plotone{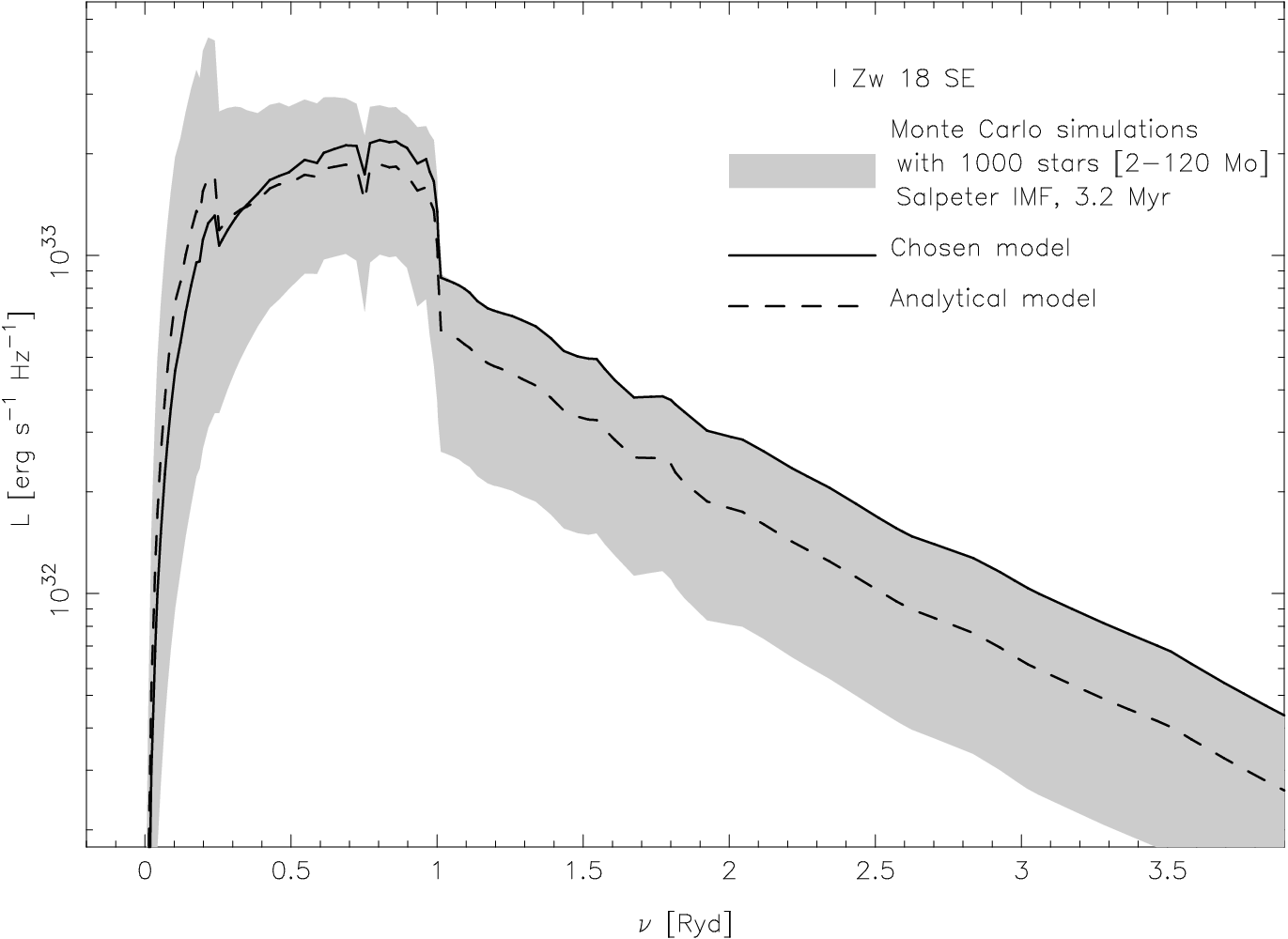}
\caption{Comparison between the analytical (dashed line) and
the Monte Carlo spectra (shaded region) of \object{I~Zw~18}.
The selected spectrum is shown as a solid line.
\label{fig:IZw18spectra}}
\end{figure}

In our models of \object{I Zw 18}
we assume an IB star-formation law; 
to obtain the spectrum, we have
performed 100 Monte Carlo simulations of clusters with 1$\times 10^3$ stars,
yielding a mean mass of 5.8 $\times 10^3$ M$_\odot$.
The complete set of simulations is shown in Figure~\ref{fig:IZw18spectra}.  
In the figure we also show for comparison the results of the analytical simulation
corresponding to the same parameters. 
As expected, the analytical simulation gives an intermediate
value with respect to the ensemble of Monte Carlo simulations. 
Since the ionizing flux in \object{I~Zw~18} is quite hard,
and since the probability of obtaining a cluster in
the upper tail of the Monte Carlo distribution is small
but appreciable,
we have chosen for our modelization one of the harder spectra.

Our best-fit model for \object{I Zw 18} 
has been obtained with a $t=3.2$ Myr spectrum,
which is shown with a bold line in Figure~\ref{fig:IZw18spectra}
(note that the figure only represents the stellar spectrum
without the nebular contribution, which could cancel out
the Balmer jump);
this spectrum gave the best results of
several alternative spectra that have also been tried.
In analogy to the case of \object{SBS~0335--052},
this model has been named model A.
The nebular parameters for the model A of \object{I~Zw~18}
are the following:
$Z_{gas} = 0.0005$;
Gaussian density law with $N_{\rm e}^{\rm max} = 150$ cm$^{-3}$, $N_{\rm e}^{\rm min} = 20$ cm$^{-3}$,
and $r_0 = 20$ pc; $\epsilon = 0.15$; $cf = 0.55$;
the chemical composition described above;
and a Str\"omgren radius $R_{\rm out} = 77$ pc.
Table~\ref{tab:IZw18comparison} compares the predicted nebular spectrum 
to the observational constraints.
The observed and predicted $T({\rm O~{\scriptstyle{II}}})$
and $T({\rm O~{\scriptstyle{III}}})$ values are also listed.

\subsection{Collisional effects in \object{I~Zw~18}}\label{sec:IZw18collisions}

Table~\ref{tab:IZw18comparison} compares the
observed to the predicted temperatures for \object{I~Zw~18};
the observed $T({\rm O~{\scriptstyle{II}}})$ and $T({\rm O~{\scriptstyle{III}}})$ values have
been taken from the paper by \citet{Ial99}.
The predicted temperatures are smaller than the observed ones,
as it happens with \object{SBS~0335--052},
but the differences in this case are far more modest,
amounting to a maximum of 250 K for the case of $T({\rm O~{\scriptstyle{II}}})$.
Additionally, \object{I~Zw~18} is somewhat colder than \object{SBS~0335--052},
so the effect of collisions on hydrogen lines is expected to be smaller.
Of bigger concern is the failure to fit the intensities
of the low-ionization lines, such as the [O{\sc~ii}] and the [S{\sc~ii}] lines.
However, in this case we have less stringent observational
constraints than in the case of \object{SBS~0335--052},
and it is not possible to obtain a better model.

The effect of temperature on the predicted collisions
was estimated following the same procedure
as in the case of \object{SBS~0335--052}:
a new model was computed, identical to model A in all respects
with the exception of metallicity, which is set to 1/10 of the
value assumed for model A. A lower scaling factor is sufficient
in this case, because the difference between the model's 
predictions and the observed temperatures are more modest
than in \object{SBS~0335--052}.
As a matter of fact, the new model -- hereinafter model B,
analogously to the nomenclature adopted for the models of \object{SBS~0335--052} -- 
turns out to be substantially hotter than \object{I~Zw~18}.
All the other predicted quantities are essentially the same
as in model A, with the only obvious exceptions of the
metal line intensities and the temperature-related quantities.
The predicted temperatures and other quantities 
are listed in Table~\ref{tab:IZw18predicted}.

Table~\ref{tab:IZw18collisions} compares the relative contribution
of collisions to the total H$\alpha$ and H$\beta$ intensities
for models A and B,
and lists the values of the derived $C({\rm H}\beta)^{\rm col}$.
The galactic reddening towards \object{I~Zw~18} is $0\,{\rm mag}\le E(B-V)\le 0.03\,{\rm mag}$
according to \citet{BH82}, and $E(B-V) = 0.032\,{\rm mag}$
according to \citet{SFD98},
yielding $C({\rm H}\beta)^{\rm gal}$ values in the range (0.00 -- 0.05) dex.
The total observed reddening is $C({\rm H}\beta)^{\rm obs} = 0.015$ dex 
according to \citet{Ial99},
and $\langle C({\rm H}\beta)^{\rm obs} \rangle = 0.10$ dex
according to \citet{Cal02} (Section~\ref{sec:IZw18CHb}).
If we adopt 
$C({\rm H}\beta)^{\rm obs} = 0.06 \pm 0.04$ dex
and $C({\rm H}\beta)^{\rm gal} = 0.025 \pm 0.025$ dex
as estimates of the observed reddening 
and the galactic extinction,
we find that the maximum amount of collisional reddening
is $C({\rm H}\beta)^{\rm col}_{\rm max} \sim 0.035 \pm 0.045$ dex.
The upper limit set by this estimate coincides
with the value given by model B, 
and is only slightly larger than the value predicted by model A.
These predictions can therefore be used to derive
a reasonable estimate for the actual collisional enhancement
of the SE knot in \object{I~Zw~18}.
The specific values assumed will be discussed in Section~\ref{sec:effectCHb}.

\section{\object{H~29}\label{sec:H29}}

\subsection{Observational constraints}

\object{H 29}, also known as \object{I Zw 36} and \object{Mrk~209},
is the third brightest H{\sc~ii} region in the sample of
24 objects analyzed by \citet{ITL97}, who measured  
$F({\rm H}\beta) = 1.46 \times 10^{-13}$ erg s$^{-1}$ cm$^{-2}$.
The heliocentric redshift of this object is 
$z=0.000937$ \citep{dVal91}, which corresponds to a velocity
relative to the Local Group centroid $v_{LG}=349$ km s$^{-1}$;
assuming $H_0 = 72$ km s$^{-1}$ Mpc$^{-1}$,
a distance $d = 4.85$ Mpc is found.
The FWHM radius in radio frequencies is $R\sim 30$ pc \citep{VT83},
while the region observed by \citet{ITL97}
corresponds to a radius $R\sim 32$ pc
(Section~\ref{sec:lines_H29}).
According to the discussion on the observed H$\beta$ flux
in Section~\ref{sec:lines_H29},
the border of the ionized region must be somewhat bigger
than this value, which we will take as a lower limit.
No high quality data are available for \object{H~29},
for several reasons: the specific problems of
each dataset will be discussed in the following.

\subsubsection{Reddening coefficient}\label{sec:H29CHb}

According to both \citet{ITL97} and \citet{MHK99}, 
the observed reddening coefficient $C({\rm H}\beta)^{\rm obs}$ is 0.00 dex,
whereas \citet{VT83} found $C({\rm H}\beta)^{\rm obs}=0.41$ dex.
While the first of these $C({\rm H}\beta)^{\rm obs}$ values is 
probably an artefact of the observational errors
and the underlying absorption in the Balmer lines,
the second value is much too high:
the detector used by \citet{VT83} was later found to be non-linear,
and the intensities given in that paper must be corrected
using the expression $S=F^{1.07}$, where $S$ is the instrumental signal and
$F$ the actual flux \citep{TPPF89}.
After correcting the data by \citet{VT83} for non-linearity
and assuming 2\AA\ of underlying absorption in H$\delta$,
a revised value $C({\rm H}\beta)=0.22\pm 0.15$ dex is obtained.
Combining this value with those by \citet{ITL97} and \citet{MHK99},
we will adopt as a rough estimate for the reddening
$C({\rm H}\beta)=0.07\pm 0.08$ dex. 

\subsubsection{Line intensities and physical conditions}\label{sec:lines_H29}

The line intensities by \citet{ITL97} have been adopted
as observational constraints. 
The line intensities by \citet{VT83} suffer from several
limitations, particularly the non-linearity of the detector
and the large observational errors.
The more recent data by \citet{ITL97} have smaller errors,
but the slit they used ($2''\times 3.5''$)
does not cover the whole object.
Comparing the H$\beta$ fluxes measured by \citet{ITL97}
and \citet{VT83} (who used a circular slit with $6.1''$ diameter),
and bearing in mind the uncertainties affecting the older data,
it can be estimated that \citet{ITL97} collected approximately 
90 percent of the total H$\beta$ flux emitted;
thus, a reasonable estimate of the total dereddened H$\beta$ intensity,
based on the $F({\rm H}\beta)$ value by \citet{ITL97}
and the reddening coefficient derived above and
corrected upwards by 10 percent, yields 
$I({\rm H}\beta) =  (1.9 \pm 0.3)\times 10^{-13}$ erg s$^{-1}$ cm$^{-2}$.
The rms density derived from $I({\rm H}\beta)$ 
is $N_{\rm e}({\rm rms}) = 45 {\rm~cm}^{-3}$.
The forbidden-line density derived from the [S{\sc~ii}] 6716/6731 ratio
by either \citet{VT83} or \citet{ITL97} is 
$N_{\rm e}({\rm rms}) \sim 100 {\rm~cm}^{-3}$,
implying an average filling factor $\epsilon \grsim 0.1$.
Nebular He{\sc~ii} $\lambda\,4686$ emission
in \object{H 29} was observed by \citet{ITL97}, 
with $I(\lambda\,4686)/I({\rm H}\beta) = 0.012$;
this line could be produced by WR stars, 
whose presence is revealed by a broad stellar feature around 4650 \AA.
The modeling procedure is described in the next section.

\subsection{The models of \object{H~29}}

\begin{figure}
\plotone{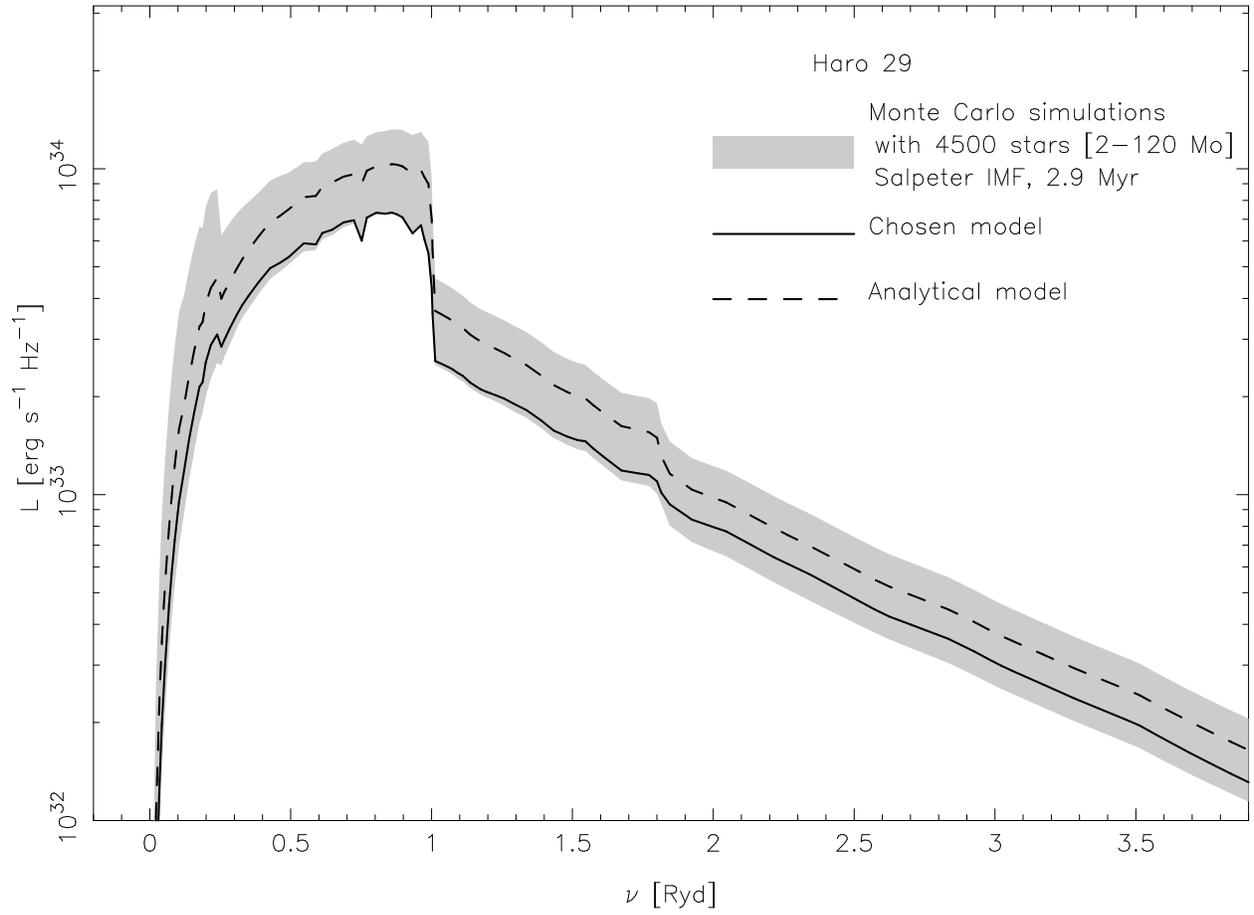}
\caption{Comparison between the analytical (dashed line) and
the Monte Carlo spectra (shaded region) of \object{H~29}.
The selected spectrum is shown as a solid line.
\label{fig:H29spectra}}
\end{figure}

We computed for \object{H~29} spherically symmetric
photoionization models with $R_S \grsim 30$ pc.
The ionizing spectra were computed assuming an IB,
and performing a sample of 100 Monte Carlo simulations with 4.5$\times 10^3$ stars;
the mean mass of the clusters turned out to be
2.65 $\times 10^4$ M$_\odot$.  
The complete set of simulations is shown in Figure~\ref{fig:H29spectra}.  
In the same figure we also show for comparison the results
of analytical simulations. 

In the models we adopted 1/10 solar abundances for all the heavy elements,
with the exception of C, N, and O, for which
we adopted abundances 1/35, 1/46, and 1/11 solar respectively,
corresponding to 12 + Log X/H = 7.01, 6.31, and 7.81.
The N and O abundances are taken from \citet{ITL97}.
For carbon there are no good determinations in \object{H~29},
so we adopted a C/O value typical of irregular galaxies
in this metallicity range \citep{Gal95,Gal97}.
The helium abundance in the gas is He/H = 0.078 by number. 

The observed total H$\beta$ intensity corresponds
to a lower limit for the rate of ionizing photons given by
$Q({\rm H}^0) \ge 1.1 \times 10^{51}$ s$^{-1}$;
on the other hand, \citet{VT83} find, by means of radio observations,
$Q({\rm H}^0) = 1.9 \times 10^{51}$ s$^{-1}$. Probably the radio
observations include outer regions of H29 not included in the H$\beta$
measurements; therefore
the comparison between these two figures gives 
$cf \le 0.58$. In the modeling we assumed $cf = 0.50$.

With these parameters, several models were computed,
varying the density law (either radially constant, or Gaussian: 
see Section~\ref{sec:SBS0335}),
coupled to the filling factor to reproduce the observed radius.
The best-fit model follows a Gaussian density law with 
$N_{\rm e}^{\rm max} = 250$ cm$^{-3}$, $N_{\rm e}^{\rm min} = 40$ cm$^{-3}$,
and $r_0 = 15$ pc; $\epsilon = 0.25$; $cf = 0.50$;
the chemical composition described above;
and a Str\"omgren radius $R_{\rm out} = 55$ pc.
This model will be called model A for consistency with the nomenclature adopted
in the previous cases; however, no enhanced-temperature alternative
models have been computed in this case, as will be explained in the following.
The predictions of this model will be presented without applying any
aperture correction, since we assume that no significant portion of the region
falls outside the slit.
Table~\ref{tab:H29comparison} compares the predicted spectrum to the 
observational constraints, 
and Table~\ref{tab:H29predicted} contains various predicted quantities
for model A.

\subsection{Collisional effects in \object{H~29}\label{sec:H29collisions}}

Table~\ref{tab:H29comparison} summarizes the
observed and predicted temperatures for \object{H~29};
the observed $T({\rm O~{\scriptstyle{III}}})$ temperatures have
been taken from the paper by \citet{Ial99}.
As it happens with \object{SBS~0335--052} and \object{I~Zw~18},
the predicted temperatures are smaller than the observed ones;
however, the differences in the case of \object{H~29} are far more modest,
amounting to a maximum of 650 K for the case of $T({\rm O~{\scriptstyle{III}}})$.
Additionally, \object{H~29} is much colder than \object{SBS~0335--052},
so the effects of collisions on hydrogen lines are certainly more
modest, and an underestimation of the temperature 
has smaller consequences;
for these reasons, we did not compute enhanced-temperature models
for this region.
Table~\ref{tab:H29collisions} compares the relative contribution
of collisions to the total H$\alpha$ and H$\beta$ intensity
for the best-fit model,
and lists the value of $C({\rm H}\beta)^{\rm col}$
for the best-fit model.
Although in the case of \object{H~29} we cannot build
a sophisticated model, due to the limitations in the
observational constraints available, 
we are confident that the collisional contributions estimated
by means of the model presented here are representative
of the real situation, on the following grounds:
first, the temperature of this region is quite low,
so that collisions are not very important;
second, the temperature of our model is a good
approximation of the temperature of \object{H~29};
third, \object{H~29} is small and compact, 
so that less structure than in the case of \object{SBS~0335--052} is expected,
making the need for spatially resolved observations less stringent;
fourth, the reddening coefficient of this region is very low,
so that the upper limit on collisional enhancement placed
by the observed reddening is also quite low.
The galactic reddening towards \object{H~29} is $0\,{\rm mag}\le E(B-V)\le 0.03\,{\rm mag}$
according to \citet{BH82}, and $E(B-V) = 0.015\,{\rm mag}$
according to \citet{SFD98},
implying $C({\rm H}\beta)^{\rm gal} = 0.02\pm 0.02$ dex.
After subtracting this contribution from the total observed reddening
$C({\rm H}\beta)^{\rm obs} = 0.07\pm 0.08$ dex,
we find that the maximum amount of collisional enhancement
for the ratio $I({\rm H}\alpha)/I({\rm H}\beta)$
is 4 percent, implying a maximum collisional enhancement of 1.5 percent
for $I({\rm H}\beta)$.

\section{Effect of the collisional enhancement of the Balmer lines on $Y_P$}

\citet{DK85} pointed out that, in objects with
high electronic temperatures,
the lower Balmer lines may be collisionally enhanced. 
If the collisional contribution is not subtracted out
from the total intensity of each line, 
it may result in spuriously low He/H ratios.
Since high temperatures characterize low-metallicity objects,
which are crucial for the determination of primordial helium,
the derived $Y_P$ is consequently underestimated.
\citet{DK85} estimated that the effect in \object{I~Zw~18} 
is roughly 2 percent for $I({\rm H}\alpha)_{\rm col}/I({\rm H}\alpha)_{\rm tot}$, 
and a third of that for $I({\rm H}\beta)_{\rm col}/I({\rm H}\beta)_{\rm tot}$.

The effect of collisional excitation of Balmer lines 
on the determination of $Y_P$ has been discussed by several authors 
after \citet{DK85} (\citealt{SK93,SI01}; Paper~I).
\citet{SI01} remarked that, because collisions affect more
H$\alpha$ than H$\beta$, collisions introduce an extra-reddening
in the Balmer spectrum, which should be subtracted out before
the abundance analysis is performed.
Altogether, they estimate that neglect of collisional excitation
of Balmer lines may result in an underestimate of $Y_P$ by as much as 5 percent.
In Paper~I, the effect of collisional enhancement of Balmer lines
was found to increase $Y_P$ by $+0.0028$, corresponding to 1 percent.
In the following, we will revise these results in the light of the 
improved models presented in this paper, and derive a new determination
of $Y_P$.

\subsection{Dust versus collisional effects}

{From} an H$\alpha$/H$\beta$ map of \object{I~Zw~18}
obtained from narrowband HST/WFPC2 images of the galaxy,
\citet{Cal02} find that the H$\alpha$/H$\beta$
ratio was not homogenous,  
presenting values ranging from 2.75 to 3.40.
After discarding several alternative explanations
for these enhanced ratios,
they interpret these variations 
in terms of dust patches in the galaxy. 
In particular, they argue against the possibility of 
collisional enhancement of H$\alpha$/H$\beta$ based on the following
line of reasoning:
a) the highest ratios are observed in the SW knot,
which is the one with the lower temperature;
b) the observed ratios do not seem to follow any definite trend
with the ionization parameter.

\begin{figure}
\plotone{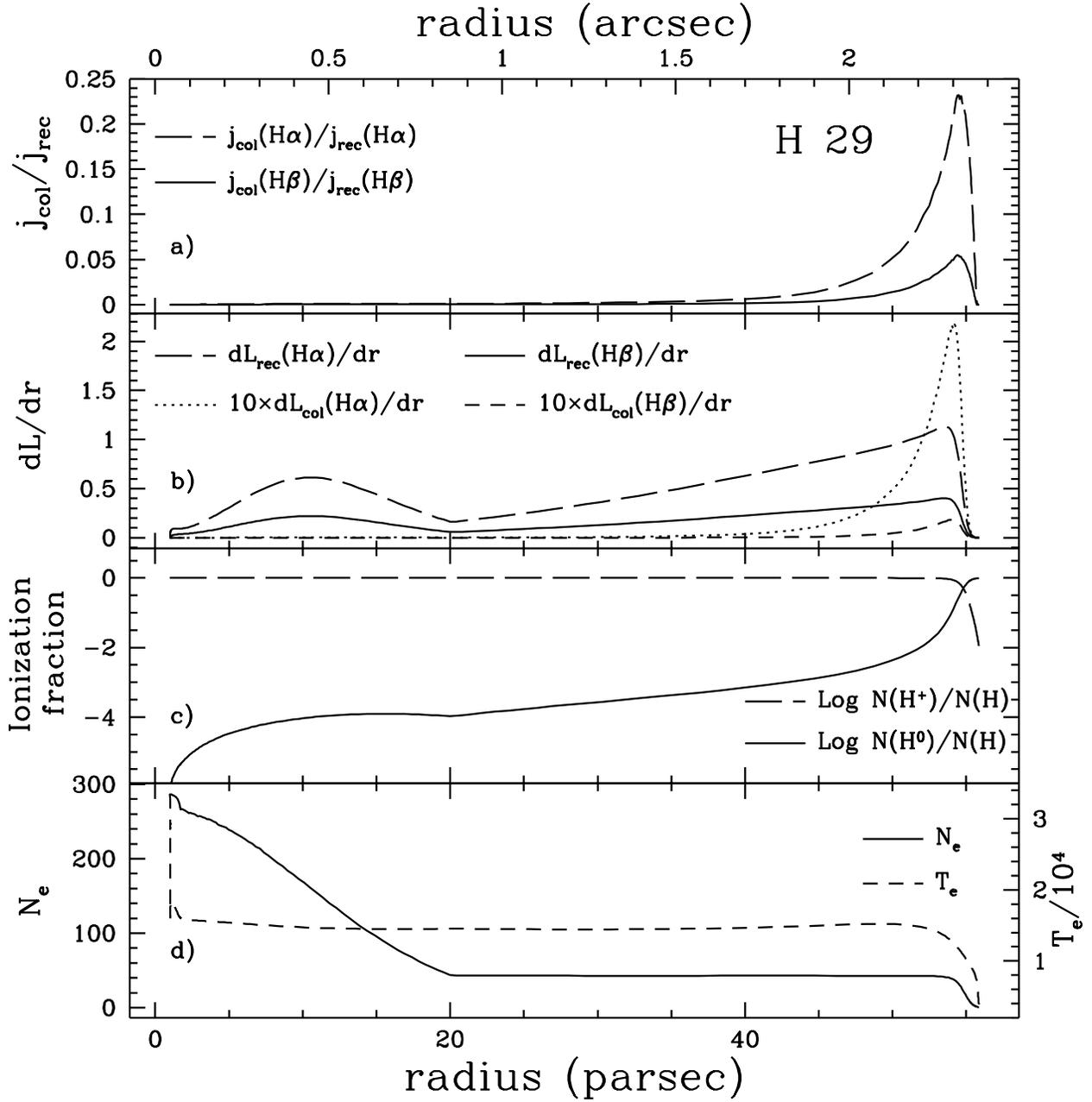}
\caption{Some predicted quantities for model A of \object{H~29}.
Panel a): radial behavior of the ratio between the collisional 
and the recombination emissivities of H$\alpha$ and H$\beta$.
Panel b): incremental contribution of each shell
to the total luminosities, with the volume factor taken into account;
units are $10^{38}$ erg sec$^{-1}$ pc$^{-1}$. 
Panel c): radial behavior of the hydrogen ionization fractions.
Panel d): radial behavior of the electron temperature and density.
\label{fig:H29profiles}}
\end{figure}

The line of reasoning expressed in the second point above
might be challenged, based on the following considerations.
In photoionization models, the collisional 
contribution reaches in general its maximum at some point halfway 
the center and the Str\"omgren radius, so the observed gradient 
depends on the exact interplay bewteen $T_e$ and the 
${\rm H}^0/{\rm H}^+$ ratio of the particular model, 
with the projection and aperture effects 
taken into account. Thus, even for a model nebula, it would be 
extremely difficulty to give a prescription for what 
the gradient in the collisional contribution is expected to be. 
One example is provided in Figure~\ref{fig:H29profiles},
which shows the behavior of various predicted quantities.
The upper panel a) contains the radial behavior of
the ratio between the collisional and the recombination emissivities
of H$\alpha$ and H$\beta$; note the abrupt rise of the curves
near the Str\"omgren radius, which is a consequence of
the increase in the fraction of neutral hydrogen outweighing
the fall in temperature.
Panel b) shows the incremental contribution of each shell
to the total luminosities, with the volume factor taken into account;
to translate this figure into an observational prediction,
one should further consider the projection effect, which in turn
depends on the aperture; nevertheless, the bumps in the plot 
makes it clear that
the dependence of luminosity on radius is something difficult
to figure out a priori.
Finally, panel c) illustrates the radial behavior of the 
hydrogen ionization fractions, and panel d) the radial 
behavior of the electron temperature and density. 
In the case of a real nebula, it is only reasonable to assume that things are 
far more complicated, and no specific shape for the gradient can be expected.
One consideration supporting this conclusion is that
the individual H{\sc~ii} regions of different stars or
small star clusters merge with one another,
a fact that makes it even more difficult to make small-scale predictions 
on the radial behavior of any quantity.
More in general, the concept of ionization parameter is tricky
and should be used with caution: one reason is that several alternative 
definitions exist, and it is not always clear which one is being used;
a second fact to consider is that most of them refer to the simplest cases 
of a central ionizing source and constant gas density, and things can
change a lot when these hypotheses are relaxed.
In short, if the concept of ionization parameter is used
as a synonim of global ionization degree, it would be better
to use the last expression, since it is more precise and univocal; 
if it refers to something else, it should be specified 
which definition is being used, and how it can be applied to the
case of a real nebula with spread-out ionizing sources and 
radially changing density and filling factor.

Nevertheless, we agree with the conclusions by \citet{Cal02},
that the main process causing H$\alpha$/H$\beta$ to depart
from its theoretical value is dust extinction,
on the following grounds:
adopting for the expected recombination ratio (H$\alpha$/H$\beta$)$_{\rm rec}$ = 2.76
as in \citet{Cal02}, 
the individual ratios measured throughout the region
differ from the expected ratio from a minimum of 0 to 
a maximum of 23 percent, averaging 8 percent.
These values yield for the total reddening coefficients
$C({\rm H}\beta)^{\rm obs}$ values from 0 to 0.27 dex.
If an average foreground extinction $C({\rm H}\beta)^{\rm gal}=0.025$ dex
is subtracted out 
\citep[Section~\ref{sec:IZw18collisions}, which agrees with the estimate by][]{Cal02},
the residual reddening amounts to values ranging from 0 to 0.24 dex,
averaging 0.07 dex.
If we were to explain such reddening in terms of pure collisional enhancement,
this effect would imply a 7 percent enhancement on average 
in the H$\alpha$ intensity
-- three times more than estimated by \citet{DK85} --,
and a 22 percent enhancement in the most extreme case 
 -- eleven times more than estimated by \citet{DK85}.
This values seem too extreme, and support the conclusion 
by \citet{Cal02} of internal dust extinction.

\subsection{Effect on the derived reddening}\label{sec:effectCHb}

Summarizing the conclusions drawn in Sections~\ref{sec:SBScollisions},
\ref{sec:IZw18collisions} and \ref{sec:H29collisions},
we will correct the observed $C({\rm H}\beta)$ values
for the three objects with our estimate of the collisional contribution.
Table~\ref{tab:CHbeta} lists the observed and collision-corrected
reddening coefficients. 

The uncertainties have been computed as follows:
for the case of \object{SBS~0335--052}, we assumed as the best
estimate for $C({\rm H}\beta)^{\rm col}$ the values predicted
with Model B, the one that best fits both the temperature and the ionization structure;
the uncertainty on $C({\rm H}\beta)^{\rm col}$ was assumed equal to the difference
between the values predicted by models A and B. 
The $C({\rm H}\beta)^{\rm obs}$ values were formally derived from the published
line fluxes.

For the case of \object{I~Zw~18},
we assumed the $C({\rm H}\beta)^{\rm col}$ value obtained with model A:
we did not adopt the higher value given by model B, as in the case of \object{SBS~0335--052},
because model B is hotter than the observations (Section~\ref{sec:IZw18collisions}), 
and because the predicted collisional enhancement would be too high
as compared to the observed reddening: see Section~\ref{sec:IZw18collisions}.
The uncertainty on $C({\rm H}\beta)^{\rm col}$ was assumed equal to the difference
between the values predicted by models A and B.
The assumed $C({\rm H}\beta)^{\rm obs}$ value is the one
inferred from the data by \citet{Cal02}, and the uncertainty by
\citet{Ial99} was assumed.

For the case of \object{H~29} we assumed the $C({\rm H}\beta)^{\rm col}$ value 
obtained with model A, and assumed the same uncertainty as in \object{I~Zw~18}.
This choice is somewhat arbitrary, but it is supported by the following considerations:
first, the observational data available to constrain models of \object{H~29}
are scarce and not high-quality;
second, the temperatures of the model are in fair agreement with the observed ones:
hence, it is reasonable to assume that the absolute uncertainty affecting
our estimation should not be too large; 
third, the temperatures (both observed and predicted) are quite low in absolute terms,
so it would be unreasonable to assume a large collisional contribution,
as it would be implied by a large uncertainty;
fourth, a larger uncertainty would contradict the assumed $C({\rm H}\beta)^{\rm obs}$ value.
As for $C({\rm H}\beta)^{\rm obs}$, the assumptions underlying the number quoted
have already been discussed in Section~\ref{sec:H29CHb}.

\section{Determination of $Y$ and $Y_P$}

In this section, we will redetermine the helium abundance in the
three H{\sc~ii} regions, taking into account the collisional contribution
to the Balmer lines. We will then show how the revised
determination affects the derived $Y_P$ value.

\subsection{Helium abundances}

As discussed in Section~\ref{sec:collisions_theorovw}, 
the abundance analysis of H{\sc~ii} regions can
suffer from a bias if the collisional enhancement of hydrogen lines
is neglected. The bias arises from two different
effects: 
the first is an overestimation of the hydrogen
recombination intensity, from which an underestimation of the
He$^+$/H$^+$ ratio directly descends;
the second effect is a
consequence of the collisional reddening of the Balmer spectrum, 
which causes an overestimation of the extinction, 
and therefore an underestimation of the intensity of the red
helium lines.  
Both effects act in the same direction, increasing the value of $Y$:
they will be illustrated quantitatively in this section,
where we present three different determinations of the helium
abundance. In the first we neglect the collisional effects 
in the hydrogen lines; 
in the second we subtract the collisional contribution 
from the observed $I({\rm H}\beta)$, but we neglect 
the collisional effects on the observed reddening;
finally, in the third we subtract the collisional contribution from both
$I({\rm H}\beta)$ and $C({\rm H}\beta)^{\rm obs}$.

In addition to the collisional contribution to the Balmer lines, to
obtain He$^+$/H$^+$ values we need a set of effective recombination
coefficients for the helium and hydrogen lines, 
an estimate of the optical depth effects for the He{\sc~i} lines,
and the contribution to the He{\sc~i} line intensities  
due to collisional excitation.
As in Paper I, we used a maximum likelihood method (MLM) 
to derive the He$^+$/H$^+$ values. 
We used the hydrogen recombination coefficients by \citet{SH95},
and the helium recombination coefficients by \citet{S96} and \citet*{BSS99}.
The collisional contribution to the He{\sc~i} lines was estimated from
\citet{SB93} and \citet{KF95}. The optical depth effects in the
triplet lines were estimated from the computations by \citet*{BSS02}.
To obtain the He$^{++}$/H$^+$ values, we used the $I$(4686)/$I$(H$\beta$)
value together with the recombination coefficients by \citet{B71}.  
The total helium abundance was derived using 
equation~(\ref{eq:ICF}) and the $icf$(He) values given by the models.

\subsubsection{Helium abundance in \object{SBS~0335--052}\label{sec:SBShelium}}

To derive He$^+$/H$^+$ in \object{SBS~0335--052} we used the following inputs
to the MLM (see Paper~I): $\tau(3889)=1.26\pm0.63$, $t^2=0.020\pm0.007$, 
and the He{\sc~i} fluxes at $\lambda\lambda$ 3889, 4026, 4471,
4922, 5876, 6678, and 7065
observed in the sum of the three centermost extractions of the slit used by \citet{Ial99}.
As in Paper I, the intensities of $\lambda\lambda$
3889, 4026, 4471, 4922, and H$\beta$ have been corrected for underlying
absorption following the prescriptions by \citet*{GDLH99}.  We did not use the
theoretical $t^2$ derived in Section~\ref{sec:SBS0335}, since the values of
Table~\ref{tab:SBSpredicted} are lower limits to the real value (see Paper I).

Table~\ref{tab:SBShelium} presents the values of the helium line fluxes, 
along with three sets of reddening-corrected line intensities.  
The $C({\rm H}\beta)$ values associated
to each intensity set are also listed.  The table also presents the $y^+$ and
$y^{++}$ determinations for each of the three cases.  Finally, the last row
presents the $Y$ values derived for each case.

\subsubsection{Helium abundance in \object{I~Zw~18}\label{sec:IZw18helium}}

Following the same procedure used with \object{SBS~0335--052}, to
derive He$^+$/H$^+$ we used the following data as
inputs for the MLM (see Paper~I): 
$\tau(3889)=0.10\pm0.05$, $t^2=0.024\pm0.007$, and
the fluxes of He{\sc~i} $\lambda\lambda$ 3889, 4026, 4471, 5876, 6678,
and 7065 observed by \citet{Ial99}. 
The intensities of $\lambda\lambda$ 3889, 4026, 4471,
and H$\beta$ have been corrected for underlying absorption according to
the work by \citet{GDLH99}. We did not use the $t^2$ 
values of Table~\ref{tab:IZw18predicted},
since they are lower limits to the real values.

Similarly to Table~\ref{tab:SBShelium}, Table~\ref{tab:IZw18helium}
presents the observed helium line fluxes from the observations of
\citet{Ial99}, along with three sets of 
reddening-corrected line intensities, 
the assumed reddening coefficients, 
and the values of $y^+$, $y^{++}$, and $Y$ determined through the MLM.

\subsubsection{Helium abundance in \object{H~29}\label{sec:H29helium}}

The He$^+$/H$^+$ value in \object{H~29} was derived using the following 
input data to the MLM:
$\tau(3889)=1.44\pm0.72$, $t^2=0.020\pm0.007$, $N_e = 175\pm 50$ cm$^{-3}$, 
and the observed fluxes of He{\sc~i} $\lambda\lambda$ 3820, 3889, 4026,
4387, 4471, 4922, 5876, 6678, 7065, and 7281. 
The intensities of $\lambda\lambda$ 3820, 3889, 4026, 4387, 4471, 4922 and
H$\beta$ have been corrected for underlying absorption, according to 
the work by \citet{GDLH99}. 
The $t^2$ listed in Table~\ref{tab:H29predicted} were not used, 
since they are lower limits to the real value.

Table~\ref{tab:H29helium} presents the observed helium line fluxes 
taken from \citet{ITL97}, along with three sets of reddening-corrected 
line intensities, the corresponding reddening coefficients,
and the $y^+$, $y^{++}$, and $Y$ values derived for each line set.

\subsection{Primordial Helium Abundance}\label{sec:Yp}

To determine $Y_P$, it is necessary to estimate for each object 
the fraction of helium produced by galactic chemical evolution. 
We will assume that
\begin{equation}
Y_P  =  Y - O \frac{\Delta Y}{\Delta O},
\label{eq:DeltaO}
\end{equation} 
where $O$ is the oxygen abundance by mass. The $\Delta O$ baseline
given by the objects in this sample is very small and consequently
produces large errors in the $\Delta Y/\Delta O$ determination:
therefore we adopted $\Delta Y/\Delta O = 3.5 \pm 0.9$, a
value based on observations and models of chemical evolution of
irregular galaxies \citep*{PPR00}.

Equation~(\ref{eq:DeltaO}) has induced researchers in this field 
to derive $Y_P$ from the analysis of
the most metal-poor extragalactic {\sc H~ii} regions known, 
particularly \object{I~Zw~18} and \object{SBS~0335--052},
in an attempt to minimize the uncertainty of the determination.
This is a valid procedure if one only considers the errors 
in the extrapolation to $O=0.00$.  
Nevertheless, the collisional effect increases with the electron
temperature, thus for very low metallicities the error due to
collisions can become very large. 
In fact, the errors due to collisions for objects with
$1/50\,{\rm Z}_\odot \la Z \la 1/25\,{\rm Z}_\odot$ 
can become larger than the errors due to the
extrapolation for objects with $1/10\,{\rm Z}_\odot \la Z \la 1/5\,{\rm Z}_\odot$;
therefore, the second set of objects allows to achieve 
a higher accuracy in the determination of $Y_P$.

From the sample made of the three objects \object{SBS~0335--052}, \object{I~Zw~18}, and
\object{H~29} we derive the following results:
(i) $Y_P = 0.2350 \pm 0.0028$, neglecting the collisional effects in the Balmer lines; 
(ii) $Y_P = 0.2380 \pm 0.0029$, subtracting the collisional contribution
from the H$\beta$ intensity, and neglecting the collisional effects on the observed reddening;
(iii) $Y_P = 0.2403 \pm 0.0030$, correcting both $I({\rm H}\beta)$ and 
$C({\rm H}\beta)$ for collisional effects. 
This is, however, a very small sample: if we consider the larger
sample presented in Paper I,
which includes \object{NGC 2363} and \object{NGC 346} in addition to the
regions studied in this work,
the $Y_P$ values obtained in the three cases become respectively
$Y_P = 0.2356 \pm 0.0018$, $Y_P = 0.2377 \pm 0.0019$, and $Y_P = 0.2391 \pm 0.0020$.

\section{Discussion}\label{sec:discussion}

There are three main procedures
to estimate the collisional contribution 
to the hydrogen line intensities in giant H{\sc~ii}
regions.
The first, which is observational, is
the measurement of $C({\rm H}\beta)^{\rm obs}$;
the difference between $C({\rm H}\beta)^{\rm obs}$ 
and $C({\rm H}\beta)^{\rm gal}$ sets an upper limit 
to the amount of collisions; no lower limit
can be found through this method.
The second method is based on the use of generic
grids of photoionization models;
although this procedure allows establishing
both a lower and an upper limit to collisional contribution, 
such limits are not necessarily more stringent than
those set with the first method. 
Furthermore, this method gives no handle on the
effect of temperature fluctuations,
which is an important source of uncertainty.
The third method is the computation of 
tailored photoionization models:
although more precise than the use of generic grids of photoionization models, 
this method is extremely time-consuming; furthermore, it suffers from the 
same bias,
since photoionization models are generally unable to reproduce
the temperatures observed in actual H{\sc~ii} regions.
This difficulty may in some way be skipped by resorting
to ingenious ways of taking the temperature bias
into account, as we did in the cases of \object{SBS~0335--052}
and \object{I~Zw~18},
but this procedure introduces an additional uncertainty
that derives from the impossibility to predict 
the spatial correlation of the temperature
to the ionization structure.

The best solution available at present is the simultaneous
use of all the methods described above. 
This approach allows to put more stringent constraints
on the amount of collisions than any of the individual 
methods taken separately; it also allows to estimate
the uncertainty attached to these estimates
in a relatively safe way, and to propagate it into 
the value of $Y_P$. 
A further refinement of this technique is at present
impossible, 
and could come only from the availability of high-quality,
spatially resolved data for a few selected objects:
spatial resolution is necessary because it allows 
to compute better photoionization models. 

{From} the point of view of the determination of $Y_P$,
a different solution to the problem
is to avoid the use of extremely low-metallicity
objects, and to concentrate instead on obtaining 
high-quality data for an adequate sample of 
moderately low-metallicity objects. 
This solution was already hinted at by \citet{P03a}. 
Although for this class of objects 
the uncertainty on $dY/dZ$ is amplified,
the uncertainty introduced by collisions is smaller,
with a global positive tradeoff.
 
\section{Summary and conclusions}\label{sec:conclusions}

In this work, we presented tailored photoionization models
for the three metal-poor extragalactic H{\sc~ii} regions
\object{SBS~0335--052}, \object{I~Zw~18}, and \object{H~29}.
In the modeling we put special care in the determination 
of the collisional contribution to the Balmer line intensities,
and gave estimates of the uncertainties affecting such determination.
An accurate determination of the collisional contribution to the Balmer lines 
requires a detailed modeling of both the temperature and the
ionization structure of the nebulae.
Although the uncertainties affecting these quantities are 
generally large,
we showed in this work that 
the collisional contribution can be estimated in a rather
precise way by means of the simultaneous use of alternative constraints.
The most powerful constraint is provided by 
the observed reddening coefficient $C({\rm H}\beta)^{\rm obs}$.
In the three objects considered in this work
the $C({\rm H}\beta)^{\rm obs}$ values are quite low,
putting a very stringent upper limit to the amount of
collisions. 
This fact has an enormous importance from the point
of view of helium determination:
these three examples suggest that
collisions are not as important as some previous analysis
had suggested \citep[e.g.][who suggested that the H$\alpha$/H$\beta$ enhancement
could reach values as high as 8 percent]{SI01};
more specifically, we can draw from this study
quantitative conclusions on the correction to
be applied to actual $Y_P$ determinations,
because our sample contains two of the objects that
have played a determining role in the past
determinations of $Y_P$.

The main results of this study can be summarized as follows:

\begin{itemize}
\item[1.] We present a set of tailored photoionization models
        for the three metal-poor extragalactic H{\sc~ii} regions
 \object{SBS~0335--052}, \object{I~Zw~18}, and \object{H~29}.
\item[2.] We computed for each model the collisional contribution
        to the H$\alpha$ and H$\beta$ intensities.
\item[3.] We discussed the different factors that could
        have an effect on our estimations, 
        and for each object determined lower and upper limits
        on the percentage of the Balmer line intensities due
        to collisions. 
\item[4.] We computed the helium abundances for each object,
        showing the effect of both neglecting and including
        Balmer line collisional effects in the analysis.
\item[5.] With the revised $Y$ values for a sample of five extremely low- and
        moderately low-metallicity H{\sc~ii} regions,
        we obtained a new determination of $Y_P$. 
        The difference with respect to the 
        case in which collisions are neglected amounts
        to an upward change of $+0.0035$, yielding $Y_P = 0.2391 \pm 0.0020$.
         
\end{itemize}

For future studies aimed at the determination of $Y_P$,
        we make the two following remarks:
\begin{itemize}
\item[1.] The use of low-metallicity objects requires 
         obtaining high-quality, spatially
         resolved, and carefully reduced data,
         which would allow to compute more sophisticated 
         photoionization models. 
\item[2.] The difficulties inherent to an accurate determination of
         collisional contributions could be avoided by directing the efforts
         towards the analysis of moderately low-metallicity H{\sc~ii} regions.
         
\end{itemize}

\acknowledgments

We would like to acknowledge the referee, Grazyna Stasi\'nska, 
for a careful reading of the paper and many useful comments.
We also thank Yuri Izotov for several constructive suggestions,
Gary Ferland for repeated help with Cloudy, and
Javier Ballesteros-Paredes and 
Guillermo Tenorio-Tagle for prompt and detailed explanations. 
This project has been partially supported by the AYA 3939-C03-01 program.
VL is supported by a Marie Curie Fellowship
of the European Community programme {\sl ``Improving Human Research Potential 
and the Socio-economic Knowledge Base''} under contract number HPMF-CT-2000-00949.
MP's work was supported in part by
grant IN 114601 from DGAPA UNAM.

\clearpage

\begin{deluxetable}{lccccccccccc}
\tabletypesize{\small}
\rotate
\tablecaption{Comparison between some observed and predicted quantities for
\object{SBS~0335--052}\tablenotemark{a}.
\label{tab:SBScomparison}}
\tablewidth{0pt}
\tablehead{
\colhead{} &&
\multicolumn{2}{c}{Center} &&
\multicolumn{2}{c}{$0''.6$} &&
\multicolumn{2}{c}{Complete slit} &&
\colhead{Complete model} \\
\cline{3-4} \cline{6-7} \cline{9-10} \cline{12-12}
\colhead{Quantity}  &&
\colhead{Observed}  &
\colhead{Predicted} &&
\colhead{Observed}  &
\colhead{Predicted} &&
\colhead{Observed}  &
\colhead{Predicted} &&
\colhead{Predicted} }
\startdata
$[$O{\sc~ii}$]$~$\lambda\,3727$/H$\beta$  && 0.253 & 0.187 && 0.256 & 0.268 && 0.279 & 0.259 && 0.319 \\
$[$O{\sc~iii}$]$~$\lambda\,4363$/H$\beta$ && 0.113 & 0.098 && 0.115 & 0.079 && 0.115 & 0.083 && 0.075 \\
He{\sc~ii}~$\lambda\,4686$/H$\beta$       && 0.031 & 0.001 && 0.032 & 0.001 && 0.032 & 0.001 && 0.001 \\
$[$O{\sc~iii}$]$~$\lambda\,5007$/H$\beta$ && 3.273 & 3.388 && 3.282 & 2.972 && 3.202 & 3.055 && 2.871 \\
$[$O{\sc~i}$]$~$\lambda\,6300$/H$\beta$   && 0.006 & 0.003 && 0.006 & 0.004 && 0.006 & 0.004 && 0.005 \\
$[$S{\sc~ii}$]$~$\lambda\,6725$/H$\beta$  && 0.035 & 0.027 && 0.036 & 0.039 && 0.039 & 0.035 && 0.046 \\
$[$O{\sc~ii}$]$~$\lambda\,7325$/H$\beta$  &&\nodata& 0.010 &&\nodata& 0.014 &&\nodata& 0.013 && 0.014 \\
$[$S{\sc~ii}$]$~$6716/6731$               && 1.071 & 1.023 && 1.063 & 1.064 && 1.121 & 1.084 && 1.175 \\
\hline
$T({\rm O~{\scriptstyle{II}}})$           && 15800 & 15500 && 15950 & 15250 && 15950 & 15250 && 15050 \\
$T({\rm O~{\scriptstyle{III}}})$          && 20400 & 18350 && 20850 & 17550 && 20950 & 17750 && 17400 \\
\hline
$R$\tablenotemark{b}                   &&  1000 &  1300 &&  1000 &  1300 &&  1000 &  1300 &&  1300 \\
Log $L({\rm H}\beta)$\tablenotemark{c}    && 40.08 & 40.12 && 39.91 & 39.89 && 40.61 & 40.61 && 41.04 \\
\enddata
\tablenotetext{a}{Comparison between some observed quantities for
\object{SBS~0335--052} and the corresponding quantities predicted according to
the multiple-shell model A, with the aperture effect included.  The H$\beta$
intensity includes the collisional contribution.  The observed [O{\sc~iii}]
temperatures, $T({\rm O~{\scriptstyle{III}}})$, are computed from the line
intensities reported by \citealt{Ial99}; the observed [O{\sc~ii}] temperatures,
 $T({\rm O~{\scriptstyle{II}}})$, have been obtained from $T({\rm
O~{\scriptstyle{III}}})$ with the analytical fit of \citealt{Ial99} 
to the models of \citealt{S90}.  The predicted $T({\rm O~{\scriptstyle{II}}})$ and $T({\rm
O~{\scriptstyle{III}}})$ values have been obtained as described in
Section~\ref{sec:output_quantities}.}
\tablenotetext{b}{In parsecs.}
\tablenotetext{c}{In erg s$^{-1}$.}
\end{deluxetable}

\clearpage

\begin{deluxetable}{lcccc}
\tablecaption{Miscellaneous predicted quantities for the multiple-shell model A of \object{SBS~0335--052}\tablenotemark{a}.\label{tab:SBSpredicted}}
\tablewidth{0pt}
\tablehead{
\colhead{Quantity}       &
\colhead{Center}         &
\colhead{$0''.6$}        &
\colhead{Complete slit}  &
\colhead{Complete model} }
\startdata
$T_0$                            & 17950 & 17150 & 17300 & 17000 \\
$T_{02}$                         & 14900 & 14750 & 14750 & 14550 \\
$T({\rm O~{\scriptstyle{II}}})$  & 15500 & 15250 & 15250 & 15050 \\
$T_{03}$                         & 18150 & 17350 & 17550 & 17250 \\
$T({\rm O~{\scriptstyle{III}}})$ & 18350 & 17550 & 17750 & 17400 \\
$t^2$                            & 0.007 & 0.007 & 0.008 & 0.008 \\
$t^2_2$                          & 0.023 & 0.020 & 0.020 & 0.019 \\
$t^2_3$                          & 0.005 & 0.004 & 0.005 & 0.005 \\
O$^+$/O                          & 0.061 & 0.092 & 0.087 & 0.108 \\
$icf({\rm He})$                  & 0.993 & 0.990 & 0.991 & 0.988 \\
\enddata
\tablenotetext{a}{This model has the following properties:
Salpeter's IMF with Monte Carlo sampling; 
$M_{up} = 120$ M$_\odot$; continuous star-formation law;
age $t = 3.0$ Myr; standard mass-loss rate; $Z_{*}=0.001$;
$Z_{gas}=0.0007$. See Section~\ref{sec:SBSmultipleshell} for further details.}
\end{deluxetable}

\clearpage

\begin{deluxetable}{lcccccccccccc}
\tabletypesize{\small}
\rotate
\tablecaption{Collisional contribution to the total Balmer intensities
for the three multiple-shell models of \object{SBS~0335--052}.
\label{tab:SBScollisions}}
\tablewidth{0pt}
\tablehead{
\colhead{} &&
\multicolumn{3}{c}{Model A} &&
\multicolumn{3}{c}{Model B} &&
\multicolumn{3}{c}{Model C} \\
\cline{3-5} \cline{7-9} \cline{11-13}
\colhead{Quantity} &&
\colhead{Center} &
\colhead{$0''.6$} &
\colhead{Slit} &&
\colhead{Center} &
\colhead{$0''.6$} &
\colhead{Slit} &&
\colhead{Center} &
\colhead{$0''.6$} &
\colhead{Slit} }
\startdata
$I({\rm H}\alpha)_{\rm col}/I({\rm H}\alpha)_{\rm tot}$ && 0.069 & 0.078 & 0.074 && 0.113 & 0.117 & 0.117 && 0.233 & 0.272 & 0.280 \\
$I({\rm H}\beta)_{\rm col}/I({\rm H}\beta)_{\rm tot}$   && 0.020 & 0.022 & 0.021 && 0.035 & 0.036 & 0.035 && 0.071 & 0.082 & 0.084 \\
\hline
$T_{03}$ && 18150 & 17350 & 17550 && 20900 & 19200 & 19100 && 20870\tablenotemark{a} & 19860\tablenotemark{a} & 20090\tablenotemark{a} \\
$C({\rm H}\beta)^{\rm col}$ && 0.07 & 0.08 & 0.07  && 0.11 & 0.11 & 0.12 && 0.25 & 0.30 & 0.31\\
\enddata
\tablenotetext{a}{These temperatures are fictitious and have been obtained by multiplying by 1.15 
the corresponding temperatures in model A (see Section~\ref{sec:SBScollisions}).}
\end{deluxetable}

\clearpage

\begin{deluxetable}{lc@{\hspace{2pt}}cccc}
\tablecaption{Comparison between some observed and predicted quantities for
\object{I~Zw~18}\tablenotemark{a}.
\label{tab:IZw18comparison}}
\tablewidth{0pt}
\tablehead{
\colhead{} &&
\multicolumn{2}{c}{Slit} && \colhead{Complete model} \\
\cline{3-4} \cline{6-6} 
\colhead{Quantity}  &&
\colhead{Observed}  &
\colhead{Predicted} &&
\colhead{Predicted} }
\startdata
$[$O{\sc~ii}$]$~$\lambda\,3727$/H$\beta$  && 0.502 &  0.202 && 0.255 \\
$[$O{\sc~iii}$]$~$\lambda\,4363$/H$\beta$ && 0.054 &  0.051 && 0.044 \\
He{\sc~ii}~$\lambda\,4686$/H$\beta$       && 0.009 &  0.001 && 0.001 \\
$[$O{\sc~iii}$]$~$\lambda\,5007$/H$\beta$ && 1.749 &  1.668 && 1.496 \\
$[$O{\sc~i}$]$~$\lambda\,6300$/H$\beta$   && 0.012 &  0.003 && 0.004 \\
$[$S{\sc~ii}$]$~$\lambda\,6725$/H$\beta$  && 0.069 &  0.031 && 0.041 \\
$[$O{\sc~ii}$]$~$\lambda\,7325$/H$\beta$  &&\nodata&  0.006 && 0.008 \\
$[$S{\sc~ii}$]$~$6716/6731$               && 1.477 &  1.390 && 1.393 \\
\hline
$T({\rm O~{\scriptstyle{II}}})$           && 15400 & 15150 && 14700 \\
$T({\rm O~{\scriptstyle{III}}})$          && 19060 & 18950 && 18650 \\
\hline
$R$\tablenotemark{b}                   && 40 -- 100  & 78 && 78 \\
Log $L({\rm H}\beta)$\tablenotemark{c}    && 38.32 &  38.32 && \nodata\\
Log $L({\rm H}\beta)$\tablenotemark{d}    && 38.49 & \nodata && 38.43 \\
\enddata
\tablenotetext{a}{Comparison between the observed line intensities relative to
H$\beta$ for the SE knot of \object{I~Zw~18} and the predictions of model A, 
with the aperture effect included.
The slit size is $1.5''\times 3.5''$.  The H$\beta$ intensity
includes the collisional contribution.  The observed $T({\rm
O~{\scriptstyle{II}}})$ and $T({\rm O~{\scriptstyle{III}}})$ are those reported
by \citealt{Ial99}. The predicted $T({\rm O~{\scriptstyle{II}}})$ and $T({\rm
O~{\scriptstyle{III}}})$ values have been obtained as described in
Section~\ref{sec:output_quantities}.}
\tablenotetext{b}{In parsecs.}
\tablenotetext{c}{In erg s$^{-1}$; observed with a $1.5''\times 3.5''$ slit (\citealt{Ial99}).}
\tablenotetext{d}{In erg s$^{-1}$; observed in the complete region (\citealt{Cal02}).}
\end{deluxetable}

\clearpage

\begin{deluxetable}{l@{\hspace{72pt}}cc}
\tablecaption{Miscellaneous predicted quantities for
models A and B of \object{I~Zw~18}.
\label{tab:IZw18predicted}}
\tablewidth{0pt}
\tablehead{
\colhead{Quantity}                  &
\colhead{Model A\tablenotemark{a}}  &
\colhead{Model B\tablenotemark{a}}  }
\startdata
$T_0$                            & 18200 & 19600 \\
$T_{02}$                         & 14650 & 14950 \\
$T({\rm O~{\scriptstyle{II}}})$  & 15150 & 15700 \\
$T_{03}$                         & 18700 & 20250 \\
$T({\rm O~{\scriptstyle{III}}})$ & 18950 & 21200 \\
$t^2$                            & 0.018 & 0.032 \\
$t^2_2$                          & 0.022 & 0.030 \\
$t^2_3$                          & 0.012 & 0.023 \\
O$^+$/O                          & 0.118 & 0.119 \\
$icf({\rm He})$                  & 0.991 & 0.991 \\
\enddata
\tablenotetext{a}{As seen through a slit of $1.5''\times 3.5''$.}
\end{deluxetable}

\clearpage

\begin{deluxetable}{l@{\hspace{72pt}}cc}
\tablecaption{Collisional contribution to the total Balmer intensities in \object{I~Zw~18}.
\label{tab:IZw18collisions}}
\tablewidth{0pt}
\tablehead{
\colhead{Quantity} &
\colhead{Model A\tablenotemark{a}} &
\colhead{Model B\tablenotemark{a}} }
\startdata
$I({\rm H}\alpha)_{\rm col}/I({\rm H}\alpha)_{\rm tot}$ & 0.060 & 0.080 \\
$I({\rm H}\beta)_{\rm col}/I({\rm H}\beta)_{\rm tot}$   & 0.017 & 0.024 \\
\hline                             
$T_{03}$                  & 18700 & 20250 \\
$C({\rm H}\beta)^{\rm col}$ & 0.06 & 0.08 \\
\enddata                                 
\tablenotetext{a}{As seen through a slit of $1.5''\times 3.5''$.}
\end{deluxetable}

\clearpage

\begin{deluxetable}{l@{\hspace{72pt}}cc}
\tablecaption{Comparison between some observed and predicted quantities of
object{H~29}\tablenotemark{a}.
\label{tab:H29comparison}}
\tablewidth{0pt}
\tablehead{
\colhead{Quantity} &
\colhead{Observed}   &
\colhead{Predicted}  }
\startdata
$[$O{\sc~ii}$]$~$\lambda\,3727$/H$\beta$          & 0.719 & 0.522 \\
$[$O{\sc~iii}$]$~$\lambda\,4363$/H$\beta$         & 0.127 & 0.097 \\
He{\sc~ii}~$\lambda\,4686$/H$\beta$               & 0.012 & 0.001 \\
$[$O{\sc~iii}$]$~$\lambda\,5007$/H$\beta$         & 5.543 & 5.203 \\
$[$O{\sc~i}$]$~$\lambda\,6300$/H$\beta$           & 0.014 & 0.009 \\
$[$S{\sc~ii}$]$~$\lambda\,6725$/H$\beta$          & 0.106 & 0.085 \\
$[$O{\sc~ii}$]$~$\lambda\,7325$/H$\beta$          &\nodata& 0.015 \\
$[$S{\sc~ii}$]$~$6716/6731$                       & 1.356 & 1.377  \\
\hline
$T({\rm O~{\scriptstyle{II}}})$\tablenotemark{b}  & 14000 & 13950 \\
$T({\rm O~{\scriptstyle{III}}})$\tablenotemark{b} & 15400 & 14750 \\
\hline
$R$\tablenotemark{c}                              & $\grsim$30 & 55 \\
Log $L({\rm H}\beta)$\tablenotemark{d}            & 38.72 & 38.70 \\
\enddata
\tablenotetext{a}{No aperture effect included. The H$\beta$ intensity includes the collisional contribution.}
\tablenotetext{b}{Both the observed $T({\rm O~{\scriptstyle{II}}})$ and
$T({\rm O~{\scriptstyle{III}}})$ are those reported by \citealt{ITL97}. The
predicted $T({\rm O~{\scriptstyle{II}}})$ and $T({\rm O~{\scriptstyle{III}}})$
values have been obtained as described in Section~\ref{sec:output_quantities}.}
\tablenotetext{c}{In parsecs.}
\tablenotetext{d}{In erg s$^{-1}$.}
\end{deluxetable}

\clearpage

\begin{deluxetable}{l@{\hspace{120pt}}cc}
\tablecaption{Miscellaneous predicted quantities for \object{H~29}.
\label{tab:H29predicted}}
\tablewidth{0pt}
\tablehead{
\colhead{Quantity}                 &
\colhead{Model A\tablenotemark{a}} }
\startdata
$T_0$                            & 14650    \\
$T_{02}$                         & 13700    \\
$T({\rm O~{\scriptstyle{II}}})$  & 13950    \\
$T_{03}$                         & 14750    \\
$T({\rm O~{\scriptstyle{III}}})$ & 14750    \\
$t^2$                            & 0.001    \\
$t^2_2$                          & 0.009    \\
$t^2_3$                          & $<$0.001 \\
O$^+$/O                          & 0.086    \\
$icf({\rm He})$                  & 0.991    \\
\enddata
\tablenotetext{a}{No aperture correction applied.}
\end{deluxetable}

\clearpage

\begin{deluxetable}{l@{\hspace{84pt}}c}
\tablecaption{Collisional contribution to the total Balmer intensities in \object{H~29}.
\label{tab:H29collisions}}
\tablewidth{0pt}
\tablehead{
\colhead{Quantity}                 &
\colhead{Model A\tablenotemark{a}} }
\startdata
$I({\rm H}\alpha)_{\rm col}/I({\rm H}\alpha)_{\rm tot}$ & 0.028 \\
$I({\rm H}\beta)_{\rm col}/I({\rm H}\beta)_{\rm tot}$   & 0.007 \\
\hline                                 
$T_{03}$                    & 13700 \\
$C({\rm H}\beta)^{\rm col}$ & 0.03 \\
\enddata
\tablenotetext{a}{No aperture correction applied.}
\end{deluxetable}

\clearpage

\begin{deluxetable}{lccccc}
\tabletypesize{\small}
\tablecaption{Observed, collisional, and collision-corrected reddening coefficients for \object{SBS~0335--052},
\object{I~Zw~18}, and \object{H~29}.\label{tab:CHbeta}}
\tablewidth{0pt}
\tablehead{
\colhead{}                 &
\multicolumn{3}{c}{\object{SBS~0335--052}} & 
\colhead{\object{I~Zw~18}} &
\colhead{\object{H~29}}    \\
\cline{2-4}
\colhead{Quantity}         &
\colhead{Center}           &
\colhead{$0''.6$}          &
\colhead{Complete slit}    &
\colhead{Complete slit}    &
\colhead{Complete model}   }
\startdata
$C({\rm H}\beta)^{\rm obs}$               & 0.23$\pm$0.02 & 0.25$\pm$0.02 & 0.26$\pm$0.02 & 0.10$\pm$0.02 & 0.07$\pm$0.08 \\
$C({\rm H}\beta)^{\rm col}$               & 0.11$\pm$0.04 & 0.11$\pm$0.03 & 0.12$\pm$0.05 & 0.06$\pm$0.02 & 0.03$\pm$0.02\\
$C({\rm H}\beta)^{\rm true}$              & 0.12$\pm$0.04 & 0.14$\pm$0.04 & 0.14$\pm$0.06 & 0.04$\pm$0.02 & 0.04$\pm$0.05\\
\enddata
\end{deluxetable}

\clearpage

\begin{deluxetable}{lcccc}
\tablecaption{Helium abundance in \object{SBS~0335--052}.
\label{tab:SBShelium}}
\tablewidth{0pt}
\tablehead{
\colhead{$\lambda$} &
\colhead{$\frac{F(\lambda)}{F({\rm H}\beta)}$\tablenotemark{a}} &
\colhead{$\frac{I(\lambda)}{I_{\rm tot}({\rm H}\beta)}$\tablenotemark{b}} &
\colhead{$\frac{I(\lambda)}{I_{\rm rec}({\rm H}\beta)}$\tablenotemark{c}} &
\colhead{$\frac{I(\lambda)}{I_{\rm rec}({\rm H}\beta)}$\tablenotemark{d}} }
\startdata
3889\tablenotemark{e}    & $0.1606\pm0.0018$ & 0.1997            & 0.2069            & 0.1958            \\
4026                     & $0.0122\pm0.0005$ & 0.0173            & 0.0179            & 0.0171            \\
4471                     & $0.0340\pm0.0006$ & 0.0382            & 0.0396            & 0.0387            \\
4686                     & $0.0292\pm0.0006$ & 0.0303            & 0.0314            & 0.0310            \\
4922                     & $0.0077\pm0.0005$ & 0.0093            & 0.0096            & 0.0097            \\
5876                     & $0.1167\pm0.0014$ & 0.1027            & 0.1064            & 0.1126            \\
6678                     & $0.0322\pm0.0006$ & 0.0263            & 0.0273            & 0.0297            \\
7065                     & $0.0425\pm0.0006$ & 0.0358            & 0.0371            & 0.0409            \\
\hline
$C({\rm H}\beta)$        & \nodata           & 0.24$\pm$0.02     & 0.24$\pm$0.02     & 0.13$\pm$0.04     \\
\hline
$\langle y^+ \rangle$    & \nodata           & 0.07635           & 0.07860           & 0.08099           \\
$\langle y^{++} \rangle$ & \nodata           & 0.00269           & 0.00279           & 0.00275           \\
$Y$                      & \nodata           & $0.2384\pm0.0048$ & $0.2438\pm0.0050$ & $0.2491\pm0.0055$ \\
\enddata
\tablenotetext{a}{Flux observed in the three centermost extractions 
of the slit used by \citealt{Ial99}, corrected for underlying absorption (see Paper I).}
\tablenotetext{b}{Reddening-corrected flux, neglecting the collisional contribution.}
\tablenotetext{c}{Reddening-corrected flux, with the collisional
contribution subtracted from $F({\rm H}\beta)$ but not from the reddening coefficient.} 
\tablenotetext{d}{Reddening-corrected flux, with the collisional
contribution subtracted from both $F({\rm H}\beta)$ and the reddening coefficient.}
\tablenotetext{e}{Sum of He{\sc~i} 3889 + H8; the value of ${I({\rm H8})}/{I({\rm H}\beta)}$ is 0.1073.}     
\end{deluxetable}

\clearpage

\begin{deluxetable}{lcccc}
\tablecaption{Helium abundance in \object{I~Zw~18}.
\label{tab:IZw18helium}}
\tablewidth{0pt}
\tablehead{
\colhead{$\lambda$} &
\colhead{$\frac{F(\lambda)}{F({\rm H}\beta)}$\tablenotemark{a}} &
\colhead{$\frac{I(\lambda)}{I_{\rm tot}({\rm H}\beta)}$\tablenotemark{b}} &
\colhead{$\frac{I(\lambda)}{I_{\rm rec}({\rm H}\beta)}$\tablenotemark{c}} &
\colhead{$\frac{I(\lambda)}{I_{\rm rec}({\rm H}\beta)}$\tablenotemark{d}} }
\startdata
3889\tablenotemark{e}    & $0.1570\pm0.0043$ & 0.2057            & 0.2093            & 0.1931            \\
4026                     & $0.0151\pm0.0036$ & 0.0210            & 0.0214            & 0.0208            \\
4471                     & $0.0352\pm0.0025$ & 0.0394            & 0.0401            & 0.0396            \\
4686                     & $0.0089\pm0.0023$ & 0.0090            & 0.0091            & 0.0091            \\
5876                     & $0.0968\pm0.0028$ & 0.0898            & 0.0914            & 0.0941            \\
6678                     & $0.0273\pm0.0019$ & 0.0246            & 0.0250            & 0.0262            \\
7065                     & $0.0249\pm0.0016$ & 0.0222            & 0.0226            & 0.0238            \\
\hline
$C({\rm H}\beta)$        & \nodata           & 0.10$\pm$0.02     & 0.10$\pm$0.02     & 0.04$\pm$0.02     \\
\hline
$\langle y^+ \rangle$    & \nodata           & 0.07552           & 0.07701           & 0.07728           \\
$\langle y^{++} \rangle$ & \nodata           & 0.00081           & 0.00082           & 0.00082           \\
$Y$                      & \nodata           & $0.2338\pm0.0072$ & $0.2373\pm0.0073$ & $0.2379\pm0.0073$ \\
\enddata
\tablenotetext{a}{Observed flux by \citealt{Ial99}, corrected for underlying
absorption (see Paper I).}
\tablenotetext{b}{Reddening-corrected flux, neglecting the collisional contribution.}
\tablenotetext{c}{Reddening-corrected flux, with the collisional
contribution subtracted from $F({\rm H}\beta)$ but not from the reddening coefficient.} 
\tablenotetext{d}{Reddening-corrected flux, with the collisional
contribution subtracted from both $F({\rm H}\beta)$ and the reddening coefficient.}
\tablenotetext{e}{Sum of He{\sc~i} 3889 + H8; the value of ${I({\rm H8})}/{I({\rm H}\beta)}$ is 0.1073.}     
\end{deluxetable}

\clearpage

\begin{deluxetable}{lcccc}
\tablecaption{Helium abundance in \object{H~29}.
\label{tab:H29helium}}
\tablewidth{0pt}
\tablehead{
\colhead{$\lambda$} &
\colhead{$\frac{F(\lambda)}{F({\rm H}\beta)}$\tablenotemark{a}} &
\colhead{$\frac{I(\lambda)}{I_{\rm tot}({\rm H}\beta)}$\tablenotemark{b}} &
\colhead{$\frac{I(\lambda)}{I_{\rm rec}({\rm H}\beta)}$\tablenotemark{c}} &
\colhead{$\frac{I(\lambda)}{I_{\rm rec}({\rm H}\beta)}$\tablenotemark{d}} }
\startdata
3820                     & $0.007\pm0.001$ & 0.010             & 0.010             & 0.010             \\
3889\tablenotemark{e}    & $0.186\pm0.001$ & 0.208             & 0.210             & 0.207             \\
4026                     & $0.016\pm0.001$ & 0.021             & 0.021             & 0.020             \\
4387                     & $0.004\pm0.001$ & 0.005             & 0.005             & 0.005             \\
4471                     & $0.037\pm0.001$ & 0.039             & 0.040             & 0.040             \\
4686                     & $0.012\pm0.001$ & 0.012             & 0.012             & 0.012             \\
4922                     & $0.009\pm0.001$ & 0.011             & 0.011             & 0.011             \\
5876                     & $0.103\pm0.001$ & 0.098             & 0.099             & 0.101             \\
6678                     & $0.029\pm0.001$ & 0.027             & 0.028             & 0.028             \\
7065                     & $0.025\pm0.001$ & 0.023             & 0.023             & 0.023             \\
7281                     & $0.005\pm0.001$ & 0.005             & 0.005             & 0.005             \\
\hline
$C({\rm H}\beta)$        & \nodata         & 0.07$\pm$0.08     & 0.07$\pm$0.08     & 0.04$\pm$0.05     \\
\hline
$\langle y^+ \rangle$    & \nodata         & 0.07776           & 0.07836           & 0.07882           \\
$\langle y^{++} \rangle$ & \nodata         & 0.00104           & 0.00104           & 0.00104           \\
$Y$                      & \nodata         & $0.2375\pm0.0040$ & $0.2389\pm0.0040$ & $0.2400\pm0.0041$ \\
\enddata
\tablenotetext{a}{Observed flux by \citealt{ITL97}, corrected for underlying
absorption (see Paper I).}
\tablenotetext{b}{Reddening-corrected flux, neglecting the collisional contribution.}
\tablenotetext{c}{Reddening-corrected flux, with the collisional
contribution subtracted from $F({\rm H}\beta)$ but not from the reddening coefficient.} 
\tablenotetext{d}{Reddening-corrected flux, with the collisional
contribution subtracted from both $F({\rm H}\beta)$ and the reddening coefficient.}
\tablenotetext{e}{Sum of He{\sc~i} 3889 + H8; the value of ${I({\rm H8})}/{I({\rm H}\beta)}$ is 0.106.}     
\end{deluxetable}

\clearpage

\clearpage

\clearpage

\clearpage

\clearpage

\clearpage

\clearpage

\clearpage

\end{document}